\documentclass[sigconf]{acmart} 
\usepackage{makecell}
\usepackage{algorithm}
\usepackage{algpseudocode}
\usepackage{amsmath}
\usepackage{algorithm,algpseudocode}
\usepackage{hyperref}
\usepackage{xspace}
\usepackage{bm}
\usepackage{gensymb}
\usepackage{enumitem}
\usepackage{balance}

\newcommand{\concept}{CHiLO\xspace}
\newcommand{\method}{ConBO\xspace}

\def\CONBO{\textsc{\method}}
\def\STANBO{\textsc{Standard BO}}
\def\MANUAL{\textsc{Manual}}

\newcommand{\xx}[1]{\colorbox{yellow}{\textcolor{black}{\small{\textbf{XX}}}}}

\definecolor{bcolor1}{rgb}     {1.0,0.0,0.0}
\definecolor{darkgreen}{rgb}     {0.0,0.5,0.0}
\definecolor{blue}{rgb}     {0,0.0,1.0}

\definecolor{red}{rgb}{1, 0, 0}
\definecolor{black}{rgb}{0, 0, 0}
\definecolor{blue}{rgb}{0, 0, 1}

\algnewcommand{\Inputs}[1]{%
  \State \textbf{Inputs:}
  \Statex \hspace*{\algorithmicindent}\parbox[t]{.8\linewidth}{\raggedright #1}
}
\algnewcommand{\Outputs}[1]{%
  \State \textbf{Outputs:}
  \Statex \hspace*{\algorithmicindent}\parbox[t]{.8\linewidth}{\raggedright #1}
}
\algnewcommand{\Initialize}[1]{%
  \State \textbf{Initialize:}
  \Statex \hspace*{\algorithmicindent}\parbox[t]{.8\linewidth}{\raggedright #1}
}

\DeclareMathOperator*{\argmax}{argmax} 

\AtBeginDocument{%
  \providecommand\BibTeX{{%
    \normalfont B\kern-0.5em{\scshape i\kern-0.25em b}\kern-0.8em\TeX}}}

\makeatletter
\def\@ACM@copyright@check@cc{}
\makeatother
\copyrightyear{2025} 
\acmYear{2025} 
\setcopyright{cc}
\setcctype{by}
\acmConference[CHI '25]{CHI Conference on Human Factors in Computing Systems}{April 26-May 1, 2025}{Yokohama, Japan}
\acmBooktitle{CHI Conference on Human Factors in Computing Systems (CHI '25), April 26-May 1, 2025, Yokohama, Japan}
\acmDOI{10.1145/3706598.3713603}
\acmISBN{979-8-4007-1394-1/25/04}

\begin{document}

\title[Continual Human-in-the-Loop Optimization]{Continual Human-in-the-Loop Optimization}

\author{Yi-Chi Liao}
\orcid{0000-0002-2670-8328}
\affiliation{%
  \institution{Department of Computer Science}
  \institution{ETH Zurich, Switzerland}
  \country{} 
}
\email{yichi.liao@inf.ethz.ch}

\author{Paul Streli}

\orcid{0000-0002-3334-7727}
\affiliation{%
  \institution{Department of Computer Science}
  \institution{ETH Zurich, Switzerland}
  \country{} 
}
\email{paul.streli@inf.ethz.ch}

\author{Zhipeng Li}
\orcid{0000-0001-6602-0176}
\affiliation{%
  \institution{Department of Computer Science}
  \institution{ETH Zurich, Switzerland}
  \country{} 
}
\email{zhipeng.li@inf.ethz.ch}

\author{Christoph Gebhardt}

\orcid{0000-0001-7162-0133}
\affiliation{%
  \institution{Department of Computer Science}
  \institution{ETH Zurich, Switzerland}
  \country{} 
}
\email{christoph.gebhardt@inf.ethz.ch}

\author{Christian Holz}
\orcid{0000-0001-9655-9519}
\affiliation{%
  \institution{Department of Computer Science}
  \institution{ETH Zurich, Switzerland}
  \country{} 
}
\email{christian.holz@inf.ethz.ch}

\renewcommand{\shortauthors}{Liao et al.}

\begin{abstract}
Optimal input settings vary across users due to differences in motor abilities and personal preferences, which are typically addressed by manual tuning or calibration. Although human-in-the-loop optimization has the potential to identify optimal settings during use, it is rarely applied due to its long optimization process. A more efficient approach would continually leverage data from previous users to accelerate optimization, exploiting shared traits while adapting to individual characteristics. We introduce the concept of Continual Human-in-the-Loop Optimization and a Bayesian optimization-based method that leverages a Bayesian-neural-network surrogate model to capture population-level characteristics while adapting to new users. We propose a generative replay strategy to mitigate catastrophic forgetting. We demonstrate our method by optimizing virtual reality keyboard parameters for text entry using direct touch, showing reduced adaptation times with a growing user base. Our method opens the door for next-generation personalized input systems that improve with accumulated experience.

\end{abstract}





\maketitle

\section{Introduction}

Due to the diverse motor abilities, preferences, and behaviors among users, optimal settings for input interactions in virtual reality (VR) and augmented reality (AR) vary significantly between individuals~\cite{9994904, 10.1145/3613904.3642354, 10.1145/3611659.3615692}. 
Today's interactive systems either commonly overlook this variability, or rely on manual user adjustment and explicit calibration procedures that can result in inefficient interactions or increased setup time.
Human-in-the-loop optimization (HiLO) presents an alternative approach that optimizes interactions based on the user's past performance with specific design parameters.

HiLO has been shown to be effective across a wide range of applications, including target selection~\cite{10024515, 10.1145/3491102.3501850}, text input \cite{9994904}, visual design~\cite{10.1145/3526113.3545664, 10.1145/3386569.3392444}, wearable devices~\cite{10.1371/journal.pone.0184054}, game development~\cite{10.1145/2858036.2858253}, and animation rendering \cite{10.5555/1921427.1921443}.
While computational optimizers help avoid exhaustive testing of every design option~\cite{7352306} and aim to identify optimal solutions with minimal user trials---thereby reducing the time users spend with suboptimal settings~\cite{10.1371/journal.pone.0184054} --- a significant barrier to HiLO's broader adoption remains its relatively low sample efficiency.
Without informed priors, HiLO often relies on initial random searches to explore the problem space, still requiring numerous trials to converge to optimal solutions.
Moreover, each user must start the optimization process from scratch~\cite{bai2023transfer}.
For instance, \citeauthor{10.1145/3491102.3501850}~\cite{10.1145/3491102.3501850} found that optimizing the transfer function for 3D selection can take 60 to 90 minutes per user.

While individual users may have distinct preferences and performance with different input settings, shared traits across the user population could be leveraged to improve the optimizer efficiency across users~\cite{10.1145/3613904.3642071}.
Ideally, optimization would become more sample-efficient as data from prior users accumulates, allowing subsequent users to benefit from earlier optimization experiences while maintaining sufficient flexibility to ensure the discovery of optimal solutions tailored to individual needs.

Thus, in this paper, we investigate the question: \emph{Can an optimizer continually learn from prior user experiences to improve its efficiency over time?}  
Despite the potential, current computational methods do not support a continually improving optimization for HiLO.
Moreover, the problem itself and the corresponding challenges have not been thoroughly formulated in the existing literature. 

One related concept is meta-Bayesian optimization \cite{feurer2018scalable, volpp2019meta, wistuba2018scalable, wistuba2021few}, which combines meta-learning and Bayesian optimization. \citet{10.1145/3613904.3642071} demonstrated the use of meta-Bayesian optimization for online HiLO, where a batch of ``prior users'' needs to go through a full optimization process from scratch, enabling the optimizer to perform more efficiently with subsequent ``end-users.'' 
However, this approach suffers from a key limitation: its computation time during deployment increases with the number of prior users, leading to scalability issues. 
Beyond a certain point, it becomes impractical, as users would experience significant delays during adaptation. 
Additionally, this method assumes that ``prior users'' can dedicate sufficient time to undergo a thorough optimization, which may not be feasible in practice.
Therefore, these methods are not suitable for direct transformation into a continual learning framework.

Another closely related concept is continual learning \cite{10444954, COSSU2021607, awasthi2019continual}, where a model improves its predictive capabilities by accumulating knowledge across different tasks (in the context of HiLO, a task is optimizing for a specific user). 
A few recent works have emerged that address continual learning within the context of optimization \cite{schur2023lifelong, kassraie2022meta}, but their target problems are limited to linear (1D) discrete bandit problems, where typical HiLO tackles interactions with continuous and multiple-dimensional parameter spaces. 
The unique challenges and the corresponding methods of continual learning for HiLO remain unaddressed and unexplored.

\begin{figure*}[t!]
    \centering
    \includegraphics[width=\linewidth]{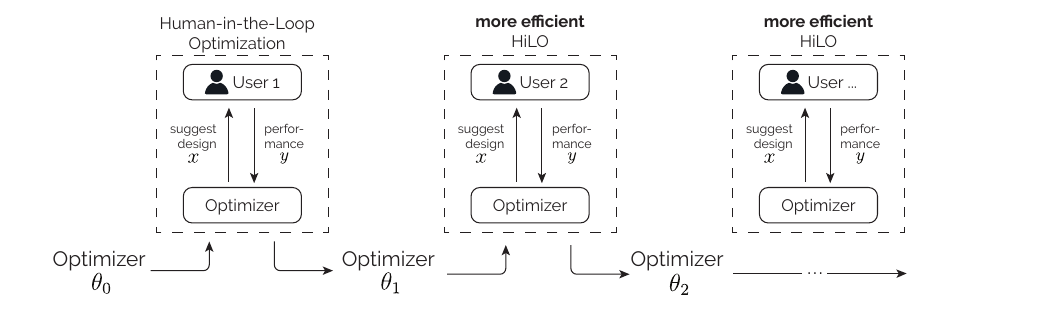}
    \caption{Illustration of \emph{Continual Human-in-the-Loop Optimization (CHiLO)}: The optimizer, parameterized by $\theta$, continuously evolves by accumulating experience from optimizations for different users. This enables more efficient adaptation to new users over time.} 
    \label{fig:concept}
\end{figure*}

To fill in the research gap, we formulate the problem and concept of \emph{Continual Human-in-the-Loop Optimization (CHiLO)} --- a computational optimizer capable of continuously improving its efficiency and performance by leveraging accumulated experience from previous users (illustrated in \autoref{fig:concept}). 
We further identify the key technical challenges that are aligned with those in continual learning \cite{10444954}, including \emph{scalability}, \emph{catastrophic forgetting}, the \emph{stability-plasticity dilemma}, and \emph{model instability due to uneven data distribution} of observations. 
These challenges guide the design principles for building \concept \space methods.
Finally, we propose our novel method, the \emph{Population-Informed Continual Bayesian Optimization (ConBO)}.
ConBO's core is a Bayesian Neural Network (BNN) \cite{mackay1995bayesian} trained on data synthesized from individual models of previous users, each representing a unique set of user characteristics. 
This method facilitates more stable and robust continual optimization by progressively integrating population-level user experiences with each new user.

We validate the efficacy and generalizability of \method using a series of standard benchmarking optimization functions, demonstrating that \method enhances optimization efficiency as more user data accumulates.
Finally, we apply \method to optimize mid-air keyboard configurations for text entry using direct touch in VR.
Our evaluations show a significant improvement in both user performance and convergence time as the number of users increases, compared to optimizing from scratch for each individual user.

In summary, we contribute
\begin{enumerate}[leftmargin=*]
    \item The concept of \emph{Continual Human-in-the-Loop Optimization}\\ (CHiLO).
    We formally define the problem, identify potential challenges, and provide design principles for future methods.
    \item Population-informed Continual Bayesian Optimization (ConBO), a first approach to CHiLO.
    ConBO trains a BNN-based optimization with data synthesized from prior users' models.
    \item A demonstration of \method's efficacy in a real-world HiLO problem.
    Here, we showcase significant performance improvements.
\end{enumerate}

\section{Related Work}

\subsection{Human-in-the-Loop Bayesian Optimization}

System settings enable the configuration of an interactive system’s parameters to optimize user interaction.
These parameters influence user performance on a given task, which can be measured through user evaluations and described using appropriate metrics~\cite{10.1145/3313831.3376262, 10.1145/3313831.3376244, 10.1145/3313831.3376735}.
Optimizing interaction for a user is thus equivalent to determining the parameters that maximize the performance metrics of interest.
Formally, we seek the optimal parameters $\bm{x}^*=\argmax_{\bm{x}\in \mathcal{X}}f\!\left(\bm{x}\right)$, where $\bm{x}$ is a design vector from the space of all possible parameters $\mathcal{X}$, dependent on the system's characteristics, and $f$ represents the relationship between system settings and user performance.

Since it remains difficult to accurately model human behavior, $f$ is generally unknown for most tasks, making optimization reliant on empirical evaluations of $f\!\left(\bm{x}'\right)$ for a given $\bm{x}'$.
Given the cost of real-world evaluations, minimizing the number of evaluations to find the global optimum is crucial.
Human-in-the-loop optimization (HiLO) addresses this by using computational optimizers to select and evaluate promising design candidates~\cite{10.1145/3313831.3376262, 10.1145/3613904.3642071, 10.1371/journal.pone.0184054, 10.1145/3526113.3545690}. 
A common approach in HiLO is Bayesian optimization~\cite{MOCKUS1975428, frazier2018tutorial, 7352306}, which guides the next configuration $\bm{x}$ to evaluate without assuming a specific functional form of $f$.
It employs two core components: a surrogate model, typically a Gaussian Process~(GP), to extrapolate a belief about unobserved points based on prior evaluations, and an acquisition function to determine the usefulness of evaluating a given point next.
Bayesian optimization has been successfully applied in various human-involved applications, such as tuning exoskeleton parameters~\cite{10.1371/journal.pone.0184054, kim2019bayesian, tucker2020human}, wearable devices~\cite{10024515,catkin2023preference, guo2019xgboost}, haptic interfaces~\cite{10024515,catkin2023preference}, input techniques~\cite{doi:10.1177/26339137241241313}, and design tools~\cite{10.5555/1921427.1921443, koyama2017sequential, 10.1145/3657643}.

However, a key limitation of Bayesian optimization is that, when initializing the surrogate model without prior knowledge, it requires multiple exploratory trials to sufficiently sample the search space~\cite{wang2022recent}.
During this exploration phase, the algorithm may evaluate severely suboptimal system configurations, which impedes the user experience. 
 This inefficiency has motivated methods that enhance Bayesian optimization by leveraging prior data through transfer learning and meta-learning.

\subsection{Transfer and Meta-Learning for Human-in-the-Loop Optimization}

Transfer learning involves pre-training a model with a task to enhance efficiency when deployed on an unseen but relevant task~\cite{weiss2016survey}. Meta-learning is a related concept, aiming to enable models to adapt quickly to new tasks by leveraging accumulated experience across multiple similar tasks~\cite{10.5555/296635, kolb2009learning, SCHWEIGHOFER20035, WANG202190}. 
While these concepts have been extensively explored in deep learning, especially in applications like few-shot recognition and reinforcement learning~\cite{pmlr-v48-santoro16,duan2016rl, bengio1990learning, 6796337,finn2017model,nichol2018reptile}, their application to optimization problems has only recently emerged~\cite{bai2023transfer}.

Bayesian optimization, with its data-driven surrogate model (typically GP), holds potential for extensions into meta-learning and transfer learning. Previous efforts have explored constructing a GP model that accumulates observations across tasks~\cite{10.5555/3042817.3042916, bonilla2007multi, NIPS2013_f33ba15e,yogatama2014efficient}. However, the primary challenge with this approach is scalability, as the GP fitting grows cubically with the number of observations; i.e., $\mathcal{O}(n^3)$ where $n$ represents the number of observations across all tasks.
Some methods have replaced components of Bayesian optimization with deep neural networks (e.g., deep kernels or deep acquisition functions) to more effectively leverage past experiences~\cite{volpp2019meta, hsieh2021reinforced,wistuba2021few}. Another approach, the weighted-sum method, constructs individual GP models for each task or user and aggregates their acquisition functions based on weighted contributions. 
This method offers better scalability~\cite{li2022transbo, 10.1007/978-3-319-46128-1_3, wistuba2018scalable}. 
For instance, a recent work, TAF$^{+}$, is applied in wrist-based input techniques~\cite{10.1145/3613904.3642071}. However, the online adaptation time of this approach increases linearly with the number of prior models (users), making it impractical for large-scale continual learning~\cite{volpp2019meta}.

Overall, current transfer and meta-Bayesian optimization methods face significant limitations that prevent their direct application to our intended goal (CHiLO), where an optimizer has to handle a sequence of users and keep improving its performance and efficiency over time. 
For example, the single GP model approach is limited by computational scalability and struggles to encode or balance users with highly diverse characteristics. 
While some works have explored using BNNs as surrogate models~\cite{springenberg2016bayesian}, these approaches are typically meant for one single problem requiring large data points to tackle, not for handling sequential tasks (users) with similar but non-identical structures. 
Continual training of neural networks on such tasks often leads to issues with model instability and catastrophic forgetting~\cite{awasthi2019continual, 10444954}.
Furthermore, weighting-based methods increase adaptation time linearly to the previously seen tasks, which limits their practical use in continual learning scenarios. Lastly, approaches like~\citet{10.1145/3613904.3642071} assume the availability of a large batch of ``prior users'' who can contribute to a long and thorough optimization process, which may not be feasible in practice. Therefore, there is a strong need for novel methods that extend Bayesian optimization into the realm of continual learning.

\subsection{Continual Learning and its Application to Optimization}

Inspired by the human ability to learn and improve through accumulating experiences across various tasks, continual learning (also known as life-long learning) is a machine learning paradigm where models are designed to learn and adapt continuously and incrementally from new data across different tasks \cite{awasthi2019continual, 10444954}. 
However, achieving this goal presents several significant challenges, including catastrophic forgetting, scalability, and maintaining model stability.
To address these challenges, various approaches have been developed. 
Regularization-based methods penalize substantial changes to critical parameters, enabling models to retain knowledge from previous tasks while learning new ones \cite{kirkpatrick2017overcoming, li2017learning}. 
Replay-based methods ensure that models revisit past data or use synthetically generated data from previous tasks to prevent the forgetting of previously acquired information \cite{rebuffi2017icarl, shin2017continual}. 
Parameter-isolation methods mitigate task interference by allocating different parts of the model's parameters to different tasks \cite{rusu2016progressive, fernando2017pathnet}. 
Additionally, meta-learning-based methods focus on continually training meta-models that can quickly adapt to new tasks by leveraging the knowledge acquired from previous tasks \cite{riemer2018learning, javed2019meta}.
Continual learning has demonstrated successful applications in domains such as computer vision \cite{yoon2017lifelong, rebuffi2017icarl}, robotics \cite{lesort2020continual, nagabandi2018deep}, natural language processing \cite{sun2019lamol, gururangan2020don}, and reinforcement learning \cite{schwarz2018progress, rolnick2019experience}.

Recent research has attempted to achieve continual learning in optimization tasks. 
One worth mentioning example is \citet{schur2023lifelong}, which introduces a method called \emph{Lifelong Bandit Optimization (LiBO)}. 
LiBO uses meta-learning to continually refine a meta-kernel function of a Gaussian Process, enabling the optimizer to accumulate experience from previous bandit problems. This method is based on meta-learning for kernels \cite{harrison2020meta, perrone2018scalable}. However, LiBO primarily addresses linear discrete bandit problems, which are significantly simpler than multi-dimensional, continuous Human-in-the-Loop Optimization (HiLO) tasks.
Moreover, a key component of LiBO is its phase of "forced exploration," making it unsuitable for our goal of minimizing random exploration for new users. 
While the lifelong bandit optimization problem is introduced in this work, its formulation and challenges are heavily focused on learning a kernel function. 
Therefore, we argue for the need to introduce the problem and challenges of Continual HiLO and to develop computational methods that extend beyond bandit problems, eliminating the requirement for forced exploration.

\subsection{Input Personalization for AR/VR Interactions}

Due to the high variance in users' motor capabilities~\cite{10.1145/3613904.3642354}, it is valuable to adapt input parameters to individual users. Traditionally, this personalization relies on explicit calibration processes, requiring users to perform specific tasks for the system to determine appropriate configurations before actual interaction begins. For example, users of wearable devices often need to perform predefined motions to allow the system to calibrate sensor ranges and adjust parameters accordingly~\cite{10.1145/2984511.2984563, 10.1145/3173574.3173755}. Similarly, gaze interactions typically involve calibration tasks where users are asked to track a series of targets~\cite{pfeuffer2013pursuit}.

More recently, input personalization has been framed as a HiLO problem. Bayesian optimization, for instance, has been applied to identify personalized transfer functions for target selection~\cite{10.1145/3491102.3501850} and to adapt keyboard dimensions for gestural typing~\cite{9994904}. However, these methods treat optimization as a more principled calibration process that is completed before the actual interaction. \citet{10.1145/3491102.3501850}, for example, requires users to undergo a 60-90 minute optimization session, while \citet{9994904} utilizes a two-stage process where users first optimize and then manually select the final design. 
Therefore, these procedures are more suitable to be a pre-interaction step rather than a real-time adaptation.

We want to emphasize that a key difference between these works and our paper is that we are not merely applying Bayesian optimization to a specific task and investigating its performance; rather, we aim to develop a framework, CHiLO, that allows the optimizer to continually improve over time and across users. This enables faster, more efficient online adaptation as the system accumulates knowledge from previous users. A related work by \citet{10.1145/3613904.3642071} employs meta-Bayesian optimization to enhance online optimization efficiency. However, as discussed earlier, their weighting-based approach is not scalable for continual learning due to challenges like computational complexity and the assumption of having a group of prior users available for extensive optimization procedures.

\section{Continual Human-in-the-Loop Optimization}

In this section, we formally define the concept of Continual Human-in-the-Loop Optimization (CHiLO), which serves as the foundation for the design principles underlying our method.

The goal of CHiLO is to optimize the parameter settings of an interactive system for a sequence of users, assuming that optimal parameters to maximize the performance metric of interest vary between individuals.
As each user arrives and undergoes the human-in-the-loop optimization, there is access to observations from prior users to progressively improve the optimizer.
The objective is to find the optimal solution for each user while minimizing the number of suboptimal trials.

\subsection{Problem Statement}
We aim to find the optimal parameters \(\bm{x}_{u}^{*} = \arg\max_{\bm{x} \in \mathcal{X}} f_u(\bm{x})\) for each user \(u \in \mathcal{U} = \{1, \cdots, N\}\), where \(\mathcal{X}\) represents the continuous multi-dimensional input space, and \(\mathcal{U}\) is the set of \(N\) users.
The function \(f_u: \mathcal{X}\rightarrow\mathbb{R}\) describes the user-dependent relationship between the input settings \(\bm{x}\) and the corresponding user performance \(f_u(\bm{x})\) according to a chosen performance metric.
Since \(f_u\) is unknown and cannot be explicitly modeled, we treat it as a black-box function during optimization.
For each user \(u\), evaluations at a given input setting \(\bm{x}\) yield an observed performance \(y = f_u(\bm{x}) + \epsilon\), where \(\epsilon \sim \mathcal{N}(0, \sigma^2)\) represents independent and identically distributed (i.i.d.) Gaussian noise, capturing the stochastic variability in user behavior.

During optimization, the optimizer, parameterized by \(\bm{\theta}\), selects a design vector \(\bm{x}_u^{t}\) at each iteration \(t\) and observes the corresponding output \(y_u^{t}\), based on the user's history of prior observations \(\mathcal{D}_u^{t-1} = \{[\bm{x}_u^0, y_u^0], [\bm{x}_u^1, y_u^1], \dots, [\bm{x}_u^{t-1}, y_u^{t-1}]\}\).
Additionally, in CHiLO, the optimizer also leverages observations from prior users \(\{\mathcal{D}_i^{T}\}_{i=1}^{u-1}\), where \(\mathcal{D}_i^{T}\) represents the history over \(T\) steps of optimization for previous users \(i\).

Thus, the design vector at iteration \(t\) is selected as
\begin{equation}
\bm{x}_u^{t} = g\left( \mathcal{D}_u^{t-1}, \{\mathcal{D}_i^{T}\}_{i=1}^{u-1} | \bm{\theta} \right),
\end{equation}
where \(g\) is the function guiding the selection process based on the current user's history and the data from prior users, with \(\bm{\theta}\) parameterizing the optimization strategy.

We aim to propose a method with an optimal selection function \(g\) and corresponding parameters \(\bm{\theta}\) such that, for each incoming user, the optimizer generates a sequence of solutions \(\{\bm{x}_u^t\}_{t=1}^{T}\) that minimize cumulative regret. The cumulative regret \(R(N, T)\) for \(N\) users over \(T\) optimization iterations is defined as

\begin{equation}
\label{regret}
R(N, T) = \sum_{u=1}^{N} \sum_{t=1}^{T} \left[ f_u(\bm{x}_u^*) - f_u(\bm{x}_u^t) \right],
\end{equation}
where \(f_u(\bm{x}_u^*)\) represents the optimal performance for user \(u\), and \(f_u(\bm{x}_u^t)\) is the performance at iteration \(t\).

The regret quantifies the cumulative difference between the optimal performance \(f_u(\bm{x}_u^*)\) and the performance at each iteration.
Minimizing the regret requires the optimizer to efficiently estimate the user's expected performance and progressively approach the optimal solution by leveraging both the user's own observations and the knowledge gathered from previous users.

\subsection{Challenges and Design Principles of \concept}
\label{sec:challenges_of_chilo}
\concept faces several challenges akin to those encountered in continual learning.
Below, we outline the most significant challenges and propose corresponding design principles to address them.

The first challenge is \textbf{scalability across a large number of users}.
For \concept to be effective, the optimizer must accumulate and leverage user experiences over time. 
As the amount of data grows, the optimizer must scale efficiently, both in encoding increasingly large datasets and maintaining computational efficiency when applied to new users.
Many traditional optimization methods struggle with this challenge.
For example, Bayesian optimization uses surrogate models like Gaussian Processes (GPs) to fit observed data.
However, due to their high computational complexity, GPs often fail to scale well with large datasets.

The second challenge is \textbf{catastrophic forgetting}.
Since \concept involves the sequential optimization of different users with unique characteristics, the optimizer may become overly specialized to the most recent users, leading to the loss of knowledge gained from prior users.
This phenomenon, known as catastrophic forgetting, can significantly hinder consistent performance across diverse user populations.
For instance, in the case of a mid-air keyboard, some users may perform better on a smaller keyboard, while others prefer a larger one.
If the optimizer adapts too strongly to one user group after consecutive optimizations, it risks forgetting how to optimize for previously encountered user types, leading to poor performance when those users reappear.

The third challenge is the \textbf{stability-plasticity dilemma}, which closely relates to the \textbf{exploration-exploitation dilemma}.
This challenge involves balancing the use of prior knowledge (exploitation) with the need to adapt to new tasks (exploration).
In \concept, new users may have characteristics that differ significantly from previous users.
If the optimizer relies too heavily on prior data, it may fail to discover the global optimum for users with novel characteristics.
On the other hand, focusing solely on new user data can lead to inefficient exploration, overfitting, or over-specialization, ultimately reducing the optimizer's ability to generalize across a diverse user population.

The fourth challenge is \textbf{model instability due to uneven distribution of observations}.
During optimization, different users may explore different regions of the design parameter space.
If certain regions are evaluated more frequently, the optimizer will accumulate more experience in those areas, leading to better predictions. 
However, this can result in neglecting other regions, causing the model to become unstable and unreliable when predicting in underexplored areas.
This issue is especially problematic in large design spaces, where certain regions may be overlooked due to perceived lower importance or higher evaluation costs.

These challenges are interrelated.
For example, poor scalability can lead to model instability as the optimizer struggles to handle large datasets, and catastrophic forgetting might occur when the optimizer focuses too narrowly on recent tasks, leading to an imbalanced distribution of observations.

In response to these challenges, we propose the following design principles for general CHiLO methods:
\begin{enumerate}
    \item \textbf{Efficient Scalability}: Methods for \concept must efficiently scale as data and the number of tasks increases.
    This requires the ability to manage large datasets and incorporate new data without significantly increasing computational complexity. 
    \item \textbf{Knowledge Retention}: Optimizers must be capable of retaining knowledge from previous tasks while adapting to new ones, thus preventing catastrophic forgetting.
    \item \textbf{Mechanism for Balancing Stability and Plasticity}:  The methods should include mechanisms that allow the optimizer to adapt to new users without disproportionately skewing the model towards highly specific user characteristics.
    \item \textbf{Robustness Across the Parameter Space}: The methods should maintain robustness and stability across the entire design parameter space, even when certain regions are underexplored or overexplored.
\end{enumerate}

\begin{figure*}[t!]
    \centering
    \includegraphics[width=\linewidth]{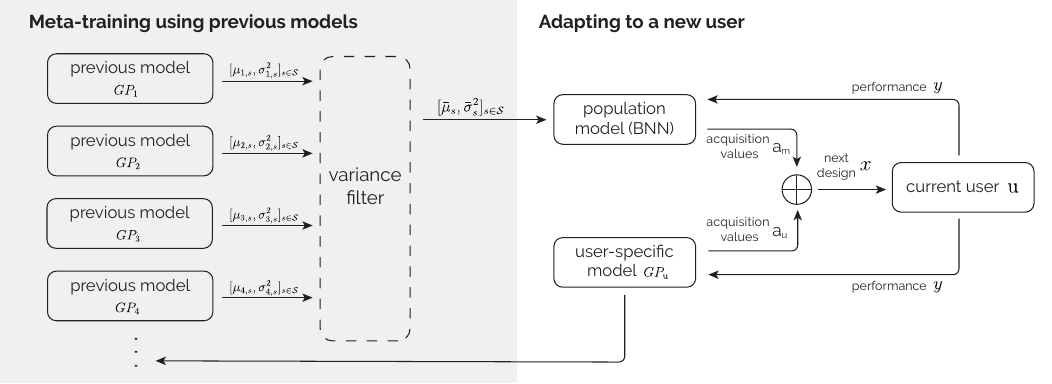}
    \caption{Illustration of the key elements and workflow of ConBO: ConBO utilizes a population model (BNN) to continually learn the population-level user characteristics. It is further trained by previous user models ($GP_{1,2,3,4...}$). We sample points across the design space and query the predicted means and variances from all GPs (\emph{meta-training}). ConBO filters out unreliable predictions, and trains the population model with the rest reliable predictions (\emph{variance filter}). When deploying on a new user $u$, the population model and a user-specific model ($GP_u$) jointly guide the optimization process (\emph{adapting to new users}). When an adaptation is complete, this user's model will be stored as a previous model.}
    \label{fig:workflow}
\end{figure*}

\section{Population-Informed Continual Bayesian Optimization (ConBO)}
\label{sec:conbo}
In line with the design principles outlined above, we propose our \textit{Population-Informed \textbf{Con}tinual \textbf{B}ayesian \textbf{O}ptimization (\textbf{ConBO})}, a novel approach that leverages population-level user characteristics to enable more stable and robust continual optimization. 
Our method is built on the foundation of Bayesian optimization (BO), which is not only a widely used method for HiLO problems \cite{7352306}, but also a powerful tool for constructing surrogate models from observed data during the optimization process.
This data-driven nature makes it well-suited for continual learning across users, as it allows for learning from data beyond just one user.
In BO, the surrogate model, typically a Gaussian Process, estimates the mean \(\mu(\bm{x})\) and variance \(\sigma^2(\bm{x})\) of a Gaussian distribution over the output of the objective function for any given design decision \(\bm{x}\), $q\left(f(\bm{x}) | \bm{x}, \mathcal{D}\right) = \mathcal{N}\left(\mu(\bm{x}), \sigma^2(\bm{x})\right)$.
The acquisition function \(a(\bm{x})\) uses these predictions to compute acquisition values, which indicate the "worth" of each design \(\bm{x}\).
The design with the highest acquisition value \(\bm{x}' = \arg\max_{\bm{x} \in \mathcal{X}} a(\bm{x})\) is selected for user evaluation, resulting in a corresponding performance measure \(y'\). This pair \((\bm{x}', y')\) is then used to update the surrogate model, refining its predictions in subsequent iterations.
As more data points \((\bm{x}, y)\) are collected, the surrogate model becomes increasingly accurate, leading to better estimates of the true function \(f(\bm{x})\), and thus guiding the optimization process more effectively.

Following the BO framework, \method relies on a surrogate model and employs a standard acquisition function.
However, in ConBO, instead of being fitted solely to a specific user's data, the surrogate model encapsulates the accumulated knowledge from all prior users, resulting in a \textit{meta-surrogate model} or \textit{population model}.
This population model facilitates more efficient optimization for new users by leveraging insights from previous user interactions, allowing for informed and effective decision-making.

Here, we outline the key components and workflow of ConBO.

\subsection{Working Principles of ConBO}

To enable \textbf{scalability}, we employ a \textit{BNN} as the population model instead of the typical GP. 
The BNN is better suited for large-scale problems, as it can predict both the mean and variance of user performance for any design while supporting BO~\cite{springenberg2016bayesian, snoek2015scalable}.
Unlike GPs, which suffer from cubic computational complexity as data grows, BNNs can handle large datasets and can be explicitly optimized to provide predictions with varying uncertainty.

To ensure \textbf{knowledge retention} across users, we implement a \textit{memory-replay mechanism}, where \method retrains the population model after each user by incorporating all accumulated data.
Without this, the population model risks forgetting previous user data by adapting only to the current user's data, leading to instability.

Addressing the \textbf{stability-plasticity dilemma}, ConBO uses a mixture-model approach \cite{feurer2018scalable, wistuba2018scalable}.
The BNN-based population model and a GP model, trained on the current user's data, work together.
Initially, the population model guides the optimization based on population-level knowledge, and as iterations progress, the GP model, tailored to the current user, takes on greater weight, allowing personalized optimization without losing the ability to generalize across users.

Ensuring \textbf{robustness across the entire design space} is essential to prevent the population model from overfitting to regions with more observations.
This is achieved through a generative memory-replay mechanism, where grid-based predictions from previous users' models are selectively retained based on uncertainty levels.
This prevents overfitting or underfitting to specific regions in the search space.

\subsection{Key Components of \method}

\method consists of these key components (see \autoref{fig:workflow}): the population model, the current user-specific model (user-specific model), and a set of previously gathered models (previous models). 

\textbf{Population model:}
The population model is a BNN designed to capture and generalize population-level user characteristics, while adapting to individual users during deployment.
Unlike a typical surrogate model that is trained on data from a single user, the population model in ConBO is trained on data synthesized from all prior users. 
Specifically, it receives grid-based predictions from previous models (described below), which include both the predicted mean and variance of performance over the design space. 
These predictions are used to retrain the population model, allowing it to generalize across users. 
The population model uses a multi-layer perceptron with dropout active during inference to capture epistemic uncertainty.
During inference, multiple samples are drawn to estimate uncertainty. The model has two linear outputs: one for the mean and another for the log-variance, which accounts for aleatoric uncertainty.
The total predictive variance is approximated as the sum of the mean output variances and the variance of predicted means, while the final mean estimate is the mean of predicted means~\cite{kendall2017uncertainties}.

During adaptation for a new user, the population model takes the design parameter settings as input, and estimates the average population mean and variances at given design points. This allows the population model to initially guide the optimization process by identifying the most promising design candidates for user evaluation based on the acquisition function values (e.g., Expected Improvement) at the selected design points.
Lastly, the population model adapts to the new user during online optimization with the observed user performance, allowing for more tailored suggestions.

\textbf{Current user-specific model (user-specific model):} For each new user currently undergoing the optimization process, ConBO constructs a GP model specifically fitted to the observations collected from that user. 
We refer to this model as the \emph{current} user-specific model (i.e., user-specific model), to differentiate from \emph{previous} user-specific models, which will be introduced in the next paragraph. 
While both the population model and the user-specific model adapt to the ongoing user's data, the user-specific model does not incorporate data from prior users. This allows it to generate acquisition values highly focused on the current user's characteristics.
The acquisition values from the user-specific model are combined with those from the population model through a weighted-sum approach. 
Initially, the population model has a stronger influence due to its population-level knowledge. 
As more iterations occur and the user-specific model gains more data, the user-specific model's influence grows, allowing for a more personalized optimization process tailored to the current user.
The use of mixed models to generate acquisition values is an established approach \cite{feurer2018scalable, wistuba2018scalable, 10.1145/3613904.3642071}. Additionally, we demonstrate the performance improvement of incorporating the user-specific model alongside the population model during adaptation, rather than relying solely on the population model, in our simulated experiments (see \autoref{appendix:experiments}).

\textbf{Previous user-specific models (previous models):}
After each user completes the optimization process, their data is used to construct a GP model, which is stored in a library of previous surrogate models. These previous models generate grid-based predictions (mean and variance) across the entire design space, which are then used to retrain the population model. 
By filtering out predictions with high variance (i.e., uncertainty), ConBO ensures that only reliable information is used for population model training.

\subsection{Workflow of \method}
\label{sec:workflow}
\method operates in a sequential manner, optimizing for one new user at a time. 
\autoref{fig:workflow} gives an overview. Following the same structure, we summarize the workflow into two primary phases, \textbf{Adaptation to a New User} and \textbf{Meta-Training}, below.

1a. \textbf{Adaptation to a new user -- proposing a design candidate:} For each new user, the population model (BNN) and the user-specific model (GP) jointly generate acquisition values over the design space.
Initially, the population model is prioritized, leveraging population-level knowledge from previous users.
As the user-specific model collects more data from the new user, its influence progressively increases.
This dynamic weighting enables efficient adaptation to new users without drastically altering the population model.
After the optimization process, the user-specific model is stored as part of the previous models. 
To manage the dynamic weighting, we follow prior work \cite{10.1145/3613904.3642071} to introduce two hyperparameters: $\alpha_1$ and $\alpha_2$ to govern the influence of both models during adaptation.
    For each surrogate model \(k\), we compute its acquisition function $a_k$ as the Expected Improvement (EI),

\begin{equation}
a_k\left(\bm{x}\right)=\mathrm{EI}_k(\bm{x}) = \int_{-\infty}^{\infty} \max(0, f(\bm{x}) - f(\bm{x}^+)) q_k(f(\bm{x}) \mid \bm{x}, \mathcal{D}) \, df(\bm{x}),
\end{equation}

\noindent
where \(\bm{x}^+\) is the design vector corresponding to the best observed value \(f(\bm{x}^+)\), and \(q_k(f(\bm{x}) \mid \bm{x}, \mathcal{D})\) is the predictive distribution of surrogate model \(k\), representing the current belief about user performance at \(\bm{x}\).

The weight of the acquisition values generated by the population model at iteration \(t\), denoted as \(w_{m,t}\), is calculated as:

\begin{equation}
\label{eq:weighting}
w_{m,t}= 
\begin{cases}
    1 ,& \text{if } t \leq \alpha_1\\
    1 - (t - \alpha_1) \alpha_2,   & \text{if } \alpha_1  < t \leq \alpha_1 + \frac{1}{\alpha_2} \\
    0,   & \text{otherwise. } 
\end{cases}
\end{equation}

\noindent
The weight for the acquisition values from the user-specific model, \(w_{u,t}\), is the complement of \(w_{m,t}\), computed as \(w_{u,t} = 1 - w_{m,t}\).
The final acquisition function for a given design candidate \(\bm{x}\) at iteration \(t\) is a weighted sum of the EI values from the population model and the user-specific model, which can be denoted as: 

\begin{equation}
a_t(\bm{x}) = w_{m,t}a_{m,t}(\bm{x}) + \left(1 - w_{m,t}\right) a_{u,t}(\bm{x}),
\end{equation}

\noindent
where $a_{m,t}$ is the acquisition value generated by the population model, and $a_{u,t}(\bm{x})$ denotes the acquisition value derived from the user-specific model. 
The design candidate with the highest acquisition value ($\bm{x}^*$) will then be selected to be evaluated by the user in this iteration; i.e., $\bm{x}_{\text{next}}=\argmax_{\bm{x}\in \mathcal{X}} a_t\!\left(\bm{x}\right)$.

Note that when \method is first deployed, it lacks sufficient knowledge from prior users, requiring early participants to undergo a certain amount of random exploration.
However, as the population model evolves and gathers more information, the need for random exploration diminishes. For later users, random exploration may no longer be necessary. 
In such cases, the user-specific model is initialized with a zero-mean function (i.e., all $x$ values lead to the same mean and variance predictions), and the optimization is fully governed by the population model until observations from the current user are collected (i.e., the iteration $t$ exceeds $\alpha_1$), at which point the GP model is updated.
These users benefit from the accumulated knowledge of earlier participants, relying initially on the population model-guided optimization.

Two additional hyperparameters control the number of random explorations required for each new user: \(r_0\) (the initial number of random explorations) and \(d_r\) (the decay rate of random exploration). 
The number of random explorations for the \(u\)-th participant (count from 1) is denoted as \(r_{u}\), which is calculated as

\begin{equation}
\label{random_number}
r_{u} = \max\left(0, r_0 - (u - 1) d_r\right).    
\end{equation}

1b. \textbf{Adaptation to a new user -- model adaptation:} 
During online adaptation, the population model keeps a record of all the observed design instances and their corresponding user performances $(x, y)$. 
We use the Gaussian negative log-likelihood loss~\cite{nix1994estimating} for a probabilistic update of the predicted mean and variance at a design point $x$, based on the observed $y$ from the user evaluation, after \emph{each new observation}.
This enables the population model to adapt to the new user's performance characteristics.
Simultaneously, the user-specific model is updated with the same observed data $(x, y)$, capturing the current user-specific uncertainty distribution over the design space.  
In later iterations, the user-specific model will primarily lead the acquisition function as its weights ($a_{u,t}(x)$) increases.  
Notably, the population model adaptation is \emph{essential} for \method, particularly during the initial optimization iterations ($t \leq \alpha_1$), where the acquisition values are entirely based on the population model. 
Without updating the population model, it would repeatedly propose the same design $\bm{x}^*$ (as defined in \autoref{eq:weighting}), further limiting the user-specific model's opportunity to learn about the user's preferences due to a lack of diverse observations in the design space.  
Once the adaptation process for a user is complete, the user-specific model is stored as a prior-user model.
Then, it will contribute to future population model training.

2a. \textbf{Meta-training -- mean and variance generation:} The population model is pre-trained using a generative replay mechanism.
Sample points are generated across the design space, following both a predefined grid and randomly selected points.
The prior-user models predict the mean and variance at these points, which form the dataset for training the population model.

2b. \textbf{Meta-training -- data synthesis and variance filtering:} Predictions with variance exceeding a threshold~$\lambda$ are considered unreliable due to high uncertainty, often caused by different users exploring distinct regions of the design space. 
Unexplored areas naturally exhibit high variance, indicating this user model's lack of information in those design regions.
To ensure the quality of the training data, predictions from unexplored regions with high uncertainty are filtered out.
If no query from a specific prior-user model yields a variance below the threshold, it suggests that this particular user's performance is highly noisy and unstable across the entire design space or simply lacks enough observations. 
In such cases, predictions from this prior-user model are excluded from population model training. 
In our later user study, we set the threshold as half of the objective value range (i.e., $\lambda = 5$ for a normalized range of $[-5, 5]$).
Importantly, we did not observe any scenario where a prior-user model's predictions were completely excluded at any point in our user study, indicating the threshold was reasonable.

2c. \textbf{Meta-training -- continual update:} The population model for a user $u$ is retrained using the filtered dataset.
The training loss consists of two terms, one for the predicted mean and one for the predicted variance.
The loss function is defined as:

\begin{equation}
\label{loss}
\mathcal{L}_{u} = \frac{1}{|\mathcal{S}|} \sum_{s\in\mathcal{S}} \left[\left(\hat{\mu}_{s} - \frac{1}{|\mathcal{U}_{u}^{s}|}\sum_{i \in \mathcal{U}_{u}^{s}}\mu_{i, s}\right)^2 + \left(\hat{\sigma}_{s}^2 - \frac{1}{|\mathcal{U}_{u}^{s}|}\sum_{i \in\mathcal{U}_{u}^{s}} \sigma_{i,s}^2 \right)^2 \right]
\end{equation}

\noindent where $\mathcal{S}$ is the set of samples used for pretraining the BNN population model, $\mathcal{U}_{u}^{s}$ is the set of prior users for user $u$ with sufficient confidence about sample $s$, \(\hat{\mu}_{s}\) and $\hat{\sigma}_{s}^2$ are the predicted mean and variance for sample $s$ from the BNN, and \(\mu_{i, s}\) and $\sigma_{i,s}^2$ are the target mean and variance according to the user-specific model (GP) of prior user $i$ for sample $s$.
When encountering a new user, this process repeats, initiating a new cycle in the workflow. The detailed of the workflow is presented in \autoref{alg:PI-BNN}.

\textbf{Summary:} \method leverages a BNN as a population model to capture population-level knowledge. While the population model adapts to the new user with the online observations, its acquisition values are combined with insights from a user-specific GP to further provide user-specific suggestions.
\method adapts to new users while minimizing the need for random exploration.
Memory replay and generative data synthesis prevent catastrophic forgetting, ensuring scalability, knowledge retention, and robustness, all in alignment with the core design principles of continual optimization.

\begin{algorithm*}
\caption{\method: Population-Informed Continual Bayesian Optimization}
\label{alg:PI-BNN}
\begin{algorithmic}[1]
\Require Users $\mathcal{U}$, design space $\mathcal{X}$, population model $q_m$ (Bayesian Neural Network), user-specific models $q_u$ (Gaussian Process, including both the previous and the current user-specific models) , number of iterations $T$, variance threshold $\lambda$, acquisition function parameters $\alpha_1$ and $\alpha_2$, random sampling parameters $r_0$ and $d_r$ 
\Ensure Optimized design choice for each user $u \in \mathcal{U}$
\State Initialize population model $q_m$ 
\State Initialize user-specific model $q_u$ for each $u \in \mathcal{U}$
\For{each user $u \in \mathcal{U}$}
    \State Determine the number of random samples $r_{u}$ based on \autoref{random_number}
    \State Sample $r_{u}$ initial design points from the design space $\mathcal{X}$, these sample design points are denoted as $ \{\bm{x}_u^0, \dots, \bm{x}_u^{r_u}\} $
    \State Perform user evaluation $f_u$ at $ \{\bm{x}_u^0, \dots, \bm{x}_u^{r_u}\} $ to obtain $ \{y_u^0, \dots, y_u^{r_u}\} $
    \State \textbf{If} $r_{u} > 0$, train user-specific model $q_u$ with $\{[\bm{x}_u^0, y_u^0], \dots, [\bm{x}_u^{r_u}, y_u^{r_u}]\}$; \textbf{Else}, initialize the user-specific model $q_u$ with a standard zero-mean function.
    \For{$t = r_{u} + 1$ to $T$}
        \State Sample $N_{\text{aq}}$ points $\bm{\tilde{x}}$ over the design space $\mathcal{X}$ to form set $\mathcal{\tilde{X}}$
        \For{each point $\bm{\tilde{x}} \in \mathcal{\tilde{X}}$}
            \State \textbf{If} $t \leq \alpha_1 + \frac{1}{\alpha_2}$, compute acquisition values $a_{m,t}(\bm{\tilde{x}}_{t})$ using the population model $q_m$; \textbf{Else}, set $a_{m,t} = 0$
            \State \textbf{If} $t > \alpha_1$, compute acquisition values $a_{u,t}(\bm{\tilde{x}}_{t})$ using the user-specific model $q_u$; \textbf{Else}, set $a_{u,t} = 0$
            \State Compute the weight $w_{m,t}$ via \autoref{eq:weighting}
            \State Compute $a_t(\bm{\tilde{x}}_{t}) = w_{m,t} a_{m,t}(\bm{\tilde{x}}) + \left(1 - w_{m,t}\right) a_{u,t}(\bm{\tilde{x}})$
        \EndFor
        \State Select next design point $\bm{x}_t = \arg\max_{\bm{\tilde{x}} \in \mathcal{\tilde{X}}} a_t(\bm{\tilde{x}})$
        \State Perform user evaluation for $f_u(\bm{x}_t)$ to obtain $y_t$
        \State Update the current user-specific model $q_u$ with $(\bm{x}_t, y_t)$
        \State Update the population model $q_m$ with $(\bm{x}_t, y_t)$ using  Gaussian negative log-likelihood loss
    \EndFor
    \State Store $q_u$ in memory

\State Sample points $\bm{x}_{s}$ across $\mathcal{X}$ to form set $\mathcal{S}$
\For{each $s \in \mathcal{S}$}
    \For{each $\tilde{u} \in \{ \tilde{u} \in \mathcal{U} \mid \tilde{u} \leq u \}$}
        \State Estimate mean $\mu_{\tilde{u}, s}$ and variance $\sigma_{\tilde{u},s}^2$ according to $q_{\tilde{u}}(f(\bm{x}_{s})|\bm{x}_{s}, \mathcal{D}_{\tilde{u}})$
    \EndFor
        \State Apply variance filtering to obtain set of samples with sufficient confidence: $\mathcal{U}^{s}_{\tilde{u}} = \{\tilde{u} \in \mathcal{U} \mid \tilde{u} \leq u \land \sigma_{\tilde{u},s}^2 < \lambda\}$
    \EndFor
    \State Re-train population model $q_m$ with filtered dataset $\mathcal{D}_{\text{filt}, u+1} = \{(\mu_{\tilde{u}, s}, \sigma_{\tilde{u}, s}^2) \mid \tilde{u} \in \mathcal{U}^{s}_{\tilde{u}}\}_{\tilde{u}\leq u}^{s \in \mathcal{S}}$ via 
    \autoref{loss}

\EndFor
\State \Return Optimized design points for all users $u \in \mathcal{U}$
\end{algorithmic}
\end{algorithm*}

\subsection{Simulated Tests with Benchmark Functions}

We evaluated ConBO's effectiveness and generalizability in simulated tests using standard benchmark functions (e.g., Branin and McCormick). 
These simulated tests, detailed in \autoref{appendix:experiments}, explored various design choices for achieving CHiLO, including a range of both GP-based and BNN-based implementations. Additionally, the tests included adaptations of existing methods to enable CHiLO, variations in replay mechanisms, the inclusion or exclusion of the user-specific model, and the use or omission of a variance filter. A complete list of the compared alternatives is provided in \autoref{appendix:different_approaches}.
The results demonstrate the advantages of \method, such as using a BNN as the population model, incorporating the user-specific model, our memory-replay mechanism, and a variance filter. 
\method outperforms alternatives in adaptation accuracy while maintaining consistent online computational costs. 

Our simulated tests highlight that GP-based methods exhibit increasing computational time as the number of observations or users grows. 
Methods that aggregate predictions across multiple GPs, such as TAF \cite{10.1145/3613904.3642071}, experience linearly increasing computation times with the number of users; that is, $\mathbf{O}(m)$, where $m$ indicates the number of previous models. 
Our experiments showed that TAF's computation time surpasses that of \method from the 4th user.
Single-GP methods, on the other hand, suffer from cubically increasing computation time with the total number of observations; that is,  $\mathbf{O}(n^3)$, where $n$ indicates the number of total observations. 
For example, our synthetic results demonstrate that the computation time of single-GP methods exceeds ConBO's after approximately 350 evaluations -- a number easily surpassed when adapting across multiple users. 
Detailed illustrations of computation costs are provided in \autoref{fig:sim1}d. These findings emphasize ConBO's effectiveness in achieving continual human-in-the-loop optimization with superior computational scalability.

Finally, in \autoref{appendix:forget}, we showcase that \method's memory-replay mechanism effectively retains the experience of previously seen tasks, leading to better performance when re-optimizing the seen synthetic users.
In contrast, without any memory-replay mechanism, the population model struggles to perform well when re-optimizing for older tasks. 

\section{User Study: Mid-Air Keyboard Personalization}

We selected text input on a mid-air keyboard in VR, using direct touch with index fingers, as a representative task to showcase the benefits of \method with real-world users.
Text entry is a fundamental input task with well-established performance metrics, and VR offers a unique opportunity for interface personalization.
Optimal keyboard size and placement varies between participants due to differences in prior experience, body size, motor abilities, and age~\cite{9994904}.
Yet, virtual keyboards are typically displayed in a predetermined, standard size.
If customization is available, it often relies on manual adjustments by the user, which, in cases of unfamiliarity with the modality, can lead to suboptimal configurations.
Thus, \method offers a unique opportunity for automatic keyboard personalization and resulting performance improvements.

\subsection{Optimization Task}

Our objective is to maximize typing performance in terms of net words per minute (net WPM)~\cite{salthouse1984effects, grady2024pressurevision}, defined as:

\begin{equation}
\label{netwpm}
\text{Net WPM} = \frac{C / 5 - E}{T/60},
\end{equation}

where $E$ is the number of character errors between the target and the typed phrase, according to the Levenshtein distance~\cite{andoni2010polylogarithmic, mackenzie_2015}, $C$ is the number of typed characters, and $T$ is the text entry duration in seconds. Based on a pilot study with three participants, we determined that a reasonable range for typing performance is between 5 and 20 net WPM. This range was then linearly normalized to an objective value of $[-5, 5]$ during the optimization process.

For keyboard configuration, we optimized both the placement (in terms of distance along the user's sagittal axis) and the scaling of the keyboard size (keeping proportions constant).
The ranges for these parameters, as shown in \autoref{tab:parameters}, were determined based on a pilot test with three participants.
These parameters were normalized to a $[0, 1]$ range during the optimization process.

\begin{table*}
    \small
    \caption{Adjustable parameters for the keyboard optimization.}
    \vspace{-1em}
        \begin{tabular}{ l l l }
            \toprule
             \multicolumn{2}{l}{\textbf{Design Parameter}} & \textbf{Range} \\
             \midrule
             $x_0$: & distance from the user along the sagittal axis & $[25, 65]\text{\,cm}$\\
             $x_1$: & scaling of the virtual keyboard (width $\times$ height) & $[39 \times 10.5$, $90 \times 31.5]$ cm\\
             \bottomrule
        \end{tabular}
    \Description{Design parameterization of absolute pointing.}
    \label{tab:parameters}
\end{table*}

\subsection{Experiment Setup}
We compared our method, \CONBO\, to standard Bayesian optimization (\STANBO) and manual user adjustment (\MANUAL) as baseline adaptation conditions.
A within-subject study was conducted, with conditions counterbalanced using a Latin-square design.
Each condition was evaluated with 10 phrases, and an optimization iteration was performed after the completion of each phrase.
Typing performance, measured in net words per minute (net WPM), was recorded after each phrase as the performance metric (objective function) for optimization.

\subsubsection{Participants, Task, and Procedure}
We recruited a total of 12 participants (2 females). Their average age is 27.75 years, ranging from 24 to 38.   
The participants' heights range from 160 to 195 cm. For further details regarding their eyesight, heights, and arm lengths, please refer to Appendix \autoref{appendix:user_details}.
In each iteration, participants were asked to transcribe a target phrase using a mid-air keyboard, optimized for size and distance along the sagittal axis.
Seated in a stationary chair, participants were instructed to type the phrase by using their index finger to poke each character individually on the mid-air keyboard.
They were instructed to type ``\textit{as quickly and as accurately as possible}'' with no option for corrections.

Each participant completed 30 iterations (phrases) throughout the study, corresponding to 10 sentences per adaptation procedure.
The target phrases were randomly selected from a set of 184 phrases, drawn from the MacKenzie \& Soukoreff phrase set~\cite{mackenzie2003phrase} and constrained to a length of 28 to 32 characters.

For \CONBO\ and \STANBO\, the keyboard size and distance were dynamically adjusted by the respective optimization method.
In the manual adjustment procedure (\MANUAL), participants had the opportunity to instruct the experimenter to adjust the keyboard’s distance to their preference before starting.
They could test the keyboard settings by typing and making as many adjustments as needed until they were satisfied.
Once a configuration was chosen, it remained fixed for the entire duration of that procedure’s 10 iterations.
Before the experiment began, participants completed a training session in which they typed 10 phrases, with keyboard sizes and distances randomly varying after each iteration.
This allowed participants to familiarize themselves with mid-air typing in VR and the study interface.

\subsubsection{Interface and Apparatus}

The study interface was implemented in Unity 2022.3.40 and run on a Windows 10 desktop computer.
The user interface consisted of three main elements: a virtual canvas displaying the target sentence, a text field for the user’s input, and a mid-air keyboard.
Participants wore a Meta Oculus Pro headset, with the Meta virtual keyboard\footnote{\url{https://developer.oculus.com/blog/virtual-keyboard-meta-quest-developers/}} used for text input.
The keyboard was positioned directly in front of the user, angled at $30\degree$ below the horizon.
Its distance from the user and size was dynamically adjusted by the respective optimization methods, with the scaling factor applied uniformly across all dimensions of the keyboard.
A virtual canvas displaying the target sentence was placed $20\degree$ above the horizon, directly in front of the user's viewpoint.
For hand-tracking, we employed the Meta hand-tracking SDK\footnote{\url{https://developer.oculus.com/documentation/unity/unity-handtracking-overview/}} integrated into Unity.

\subsection{Parameter Settings of Adaptive Keyboard Optimizers}

\subsubsection{\CONBO\ configuration}
In the following section, we will detail the implementation of \method's components: (1)~the population model (BNN), (2)~the current and previous user-specific models (GP), (3)~the settings used for adapting to new users, and (4)~the settings applied during population model training.

\paragraph{1. population model Configuration:}
Our \emph{population model} is a Bayesian Neural Network consisting of three fully connected layers, each with 100 nodes in the hidden layers.
The input dimension is 2, corresponding to the two design parameters (as described in \autoref{tab:parameters}), while the output represents the prediction of the objective function (net WPM) and the estimated aleatoric variance.
Dropout layers, with a dropout rate of 0.1, are applied after the last hidden layer to introduce stochasticity and provide regularization.
The Rectified Linear Unit (ReLU) is used as the activation function.
From a 2-people pilot test, we set the training epoch as such: During population model training (Step 4 in the workflow), BNN is trained with 800 epochs. 
During online adaptation (Step 1 in the workflow), 20 epochs are run for each new observation (iteration). 

\paragraph{2. Current and Previous User-specific Models Configuration:}
For the current and prior \emph{user-specific models} in \method, we implemented Gaussian Process (GP) models, following the approaches by \citeauthor{10.1145/3613904.3642071}~\cite{10.1145/3613904.3642071} and \citeauthor{10.1145/3491102.3501850}~\cite{10.1145/3491102.3501850}.
The GPs are single-task and use a Matérn $5/2$ kernel, which models homoscedastic noise levels in the observed data\footnote{For more details, see \url{https://botorch.org/docs/models}.}. This kernel offers a flexible approach for modeling smooth functions while accounting for varying length scales in the input space.

\paragraph{3. Adaptation to Unseen Users:}
When deploying \method for new users, we gradually reduce the number of random explorations as more participants are added.
The initial number of random explorations, $r_0$, is set to 6, with a decay rate $d_r$ of 2.
Thus, the first participant experiences 6 random explorations followed by 4 optimization iterations.
The second participant experiences 4 random explorations and 6 optimization iterations, and so on.
Starting from the fourth participant, no random explorations are conducted; all suggested keyboard settings are generated based entirely on the combined acquisition function, following Expected Improvement.
We merge the EI values from both the population model and the current user-specific model using the weighting factor from ~\autoref{eq:weighting}.
Specifically, when computing the acquisition values, we sampled a $40 \times 40$ grid evenly across the 2-dimensional design space (i.e., $N_{\text{aq}ß} = 40 \times 40$ in \autoref{alg:PI-BNN}).
This fine resolution ensures better accuracy in identifying the optimal acquisition value, as this directly impacts the design point selected for optimization during interactions with unseen users.
The hyperparameters $\alpha_1 = 5$ and $\alpha_2 = 0.2$ ensure that the influence of the current user’s model begins to increase after the fifth iteration and then grows linearly.

\paragraph{4. Population model (BNN) Training:}
To train the population model, we first generate a grid of $20 \times 20$ points, evenly spaced across the entire design space, followed by 100 randomly sampled points. 
Note that BNNs generally produce smooth predictions, enabling effective interpolation between sampled points. The $20 \times 20$ grid balances computational efficiency with capturing general trends across the design space. The additional 100 randomly sampled points further diversify the training data, reducing the need for a finer grid.
For each point, we query the predicted means and variances from all previous user-specific models.
We set a variance threshold of $\lambda = 5$.
Since the normalized objective range is 10, any variance exceeding 5 (half the objective range) is considered unreliable and is excluded from further consideration.
This approach ensures that the population model's predictions remain robust and reliable for guiding future optimizations.

\subsubsection{\STANBO\ configuration}

For \STANBO\, we used a single-task GP with Matern $5/2$ kernel.
We configured the optimization with 10 restarts for the acquisition function, 1,024 restart candidates for the acquisition function optimization, and 512 Monte Carlo samples to approximate the acquisition function.

Based on insights from a pilot test with two participants, we determined that the first 6 iterations of the 10-iteration procedure would be random searches, followed by 4 iterations of optimization. As in \method, we used Expected Improvement as the acquisition function to guide the optimization process.

\subsection{Results}

\begin{figure*}[t!]
    \centering
    \includegraphics[width=\linewidth]{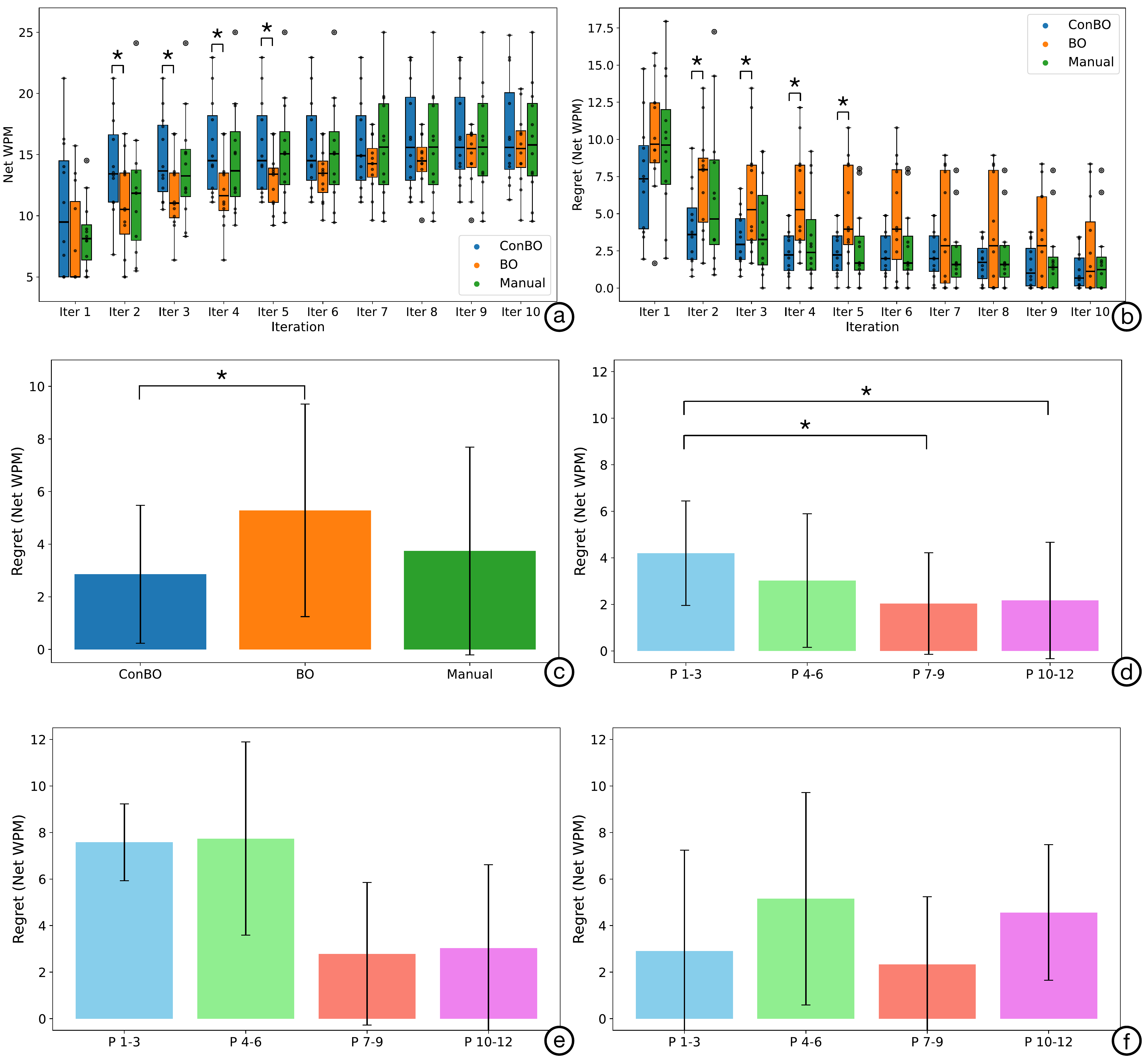}
    \caption{Results of our user study. (a) The net WPM at each iteration for three adaptation procedures; the $*$ sign indicates a significant difference found between the \method and \STANBO\ at that iteration. The scattered dots visualize each data point from individual participants. (b) The regret across all the iterations for all three procedures; the $*$ sign indicates a significant difference between the \method and \STANBO\. The scattered dots visualize each data point from individual participants. (c) The average regret for all conditions; a significant difference found \method and \STANBO\. (d) The mean regret values for each user group using \CONBO; the $*$ sign indicates a significant difference between the groups. (e) The mean regret values for each user group using \STANBO. (f) The mean regret values for each user group using \MANUAL.} 
    \label{fig:result}
\end{figure*}

To evaluate optimization performance, we analyze the maximum typing performance (net WPM) achieved up to each iteration for each user between conditions. The typing speed results are shown in \autoref{fig:result}a. We conducted a two-way repeated measures ANOVA to examine the effects of the \textit{adaptation procedures} and \textit{iteration} on performance (Net WPM, \textsc{Condition $\times$ Iteration}).

The analysis revealed no significant interaction effect between condition and iteration, $F(18, 198) = 0.57$, $p = 0.561$, indicating that the influence of condition on performance did not significantly change across iterations.
However, there was a significant main effect of condition, $F(2, 22) = 3.94$, $p = 0.039$, indicating performance differences across conditions.
Additionally, we found a highly significant main effect of iteration, $F(9, 99) = 1487.13$, $p < 0.001$, suggesting that performance significantly improved throughout the iterations.
To further explore the significant main effect of the procedures, pairwise comparisons with \textit{Bonferroni correction} were performed. A significant difference was found between \CONBO\ and \STANBO\, $t(11) = 2.97$, $p = 0.038$, indicating that \CONBO\ outperforms \STANBO\ with a moderate effect size ($\text{Hedges' g} = 0.733$). No significant differences were observed between \CONBO\ and \MANUAL\ ($t(11) = 1.04$, $p = 0.965$) or between \STANBO\ and \MANUAL\ ($t(11) = -1.53$, $p = 0.460$).
Building on the overall performance differences between procedures, we conducted one-way repeated measures ANOVA for each iteration. Significant differences were found at iterations 2, 3, 4, and 5, where \CONBO\ significantly outperformed \STANBO.

These results demonstrate that \CONBO\ outperforms \STANBO\ in terms of maximum typing performance.
Alternatively, we can assess the differences in regret throughout the optimization process (\autoref{regret}) across iterations and participants.
Regret quantifies the difference between the achieved performance and the user's optimal achievable performance.
For each user, we use the highest observed performance across all procedures and iterations as a proxy for their optimal performance.
Essentially, regret reflects the efficiency of the adaptation method, capturing the extent to which users spent in suboptimal configurations.

We visualize the regret for all conditions at each iteration in \autoref{fig:result}b. A two-way repeated measures ANOVA was conducted to examine the effects of \textit{adaptation procedures} and \textit{iteration} on regret values. We did not find a significant interaction between condition and iteration (\textsc{condition $\times$ iteration}), $F(18, 198) = 0.91$, $p = 0.561$. However, there was a significant main effect of condition, $F(2, 22) = 3.76$, $p = 0.039$, indicating that regret differed across conditions. Additionally, there was a highly significant main effect of iteration, $F(9, 99) = 46.12$, $p < 0.001$, suggesting that regret significantly changed over time.
Pairwise comparisons using \textit{Bonferroni correction} revealed significant differences between \CONBO\ and \STANBO, $t(11) = 2.97$, $p = 0.038$, with a moderate effect size ($\text{Hedges’ g} = 0.995$). No significant difference was found between \STANBO\ and \MANUAL, $t(11) = 1.53$, $p = 0.460$, or between \CONBO\ and \MANUAL, $t(11) = -1.04$, $p = 0.965$.

We further visualize the overall regret values for each condition in \autoref{fig:result}c, highlighting the differences across procedures. An additional repeated measures ANOVA was conducted and showed a significant difference between conditions with \textit{F}(2, 22) = 3.76, \textit{p} = 0.039. 
Post-hoc pairwise comparison with Bonferroni correction showed that a significant difference was found between \CONBO\ and \STANBO\ (\textit{t}(11) = 2.97, \textit{p} = 0.038).
To summarize all the users and all iterations (as per \autoref{regret}), \CONBO, \STANBO, and \MANUAL\ have total regrets of 342.79, 634.41, and 449.06 net WPMs, respectively. 
Thus, based on both net WPM and regret, we conclude that \CONBO\ outperforms \STANBO.

We analyze whether \CONBO\ enables progressively more efficient adaptation as more users are introduced. To investigate this, we divided the users into four groups based on the order in which they participated (\autoref{fig:result}d): Group 1 consists of participants 1 to 3, Group 2 includes participants 4 to 6, Group 3 includes participants 7 to 9, and Group 4 comprises participants 10 to 12.
We examine the observed regrets within each group using a one-way ANOVA.
The results showed a significant difference across the four groups, $F(3, 40) = 4.75$, $p < 0.004$.
To further explore the differences between groups, we conducted post-hoc Tukey's HSD tests.
There was a significant difference between Group 1 and Group 3, with Group 1 showing significantly higher regret than Group 3 ($\text{mean difference} = -2.17$, $p = 0.006$, 95\% CI $[-3.85, -0.48]$).
Similarly, a significant difference was found between Group 1 and Group 4, with Group 1 displaying significantly higher regret than Group 4 ($\text{mean difference} = -2.03$, $p = 0.011$, 95\% CI $[-3.72, -0.34]$).
As comparison baselines, we also plot the regret values when using \STANBO\ and \MANUAL\ grouped every 3 participants in \autoref{fig:result}d and \autoref{fig:result}e.

Finally, \autoref{appendix:user_diversity} presents additional analyses. 
Appendix \autoref{appendix:user_details} provides detailed information about the participants, while Appendix \autoref{appendix:optimal_keyboards} highlights the high diversity of optimal keyboard settings across users generated from various conditions, demonstrating ConBO's effectiveness in producing customized settings. 
Appendix \autoref{appendix:meta_model} analyzes the progressive evolution of the population model as it adapts across users.

\subsection{Findings and Discussion}

We found that \CONBO\ enables faster typing speeds with lower regret values compared to \STANBO\, proving it to be a more efficient keyboard adaptation method that benefits from the continually adapting population model.
In the first five iterations, \STANBO\ performs worse, as expected, due to the need for random searches to explore the problem space. \CONBO\ improves more rapidly over these iterations, demonstrating its ability to adapt to individuals and extend optimization beyond the population model's initial suggestions.

\MANUAL\ offers a seeming advantage by allowing participants to freely adjust keyboard settings before starting.
On average, participants made 4.92 adjustments ($std = 1.85$) before confirming their preferred setup.
Despite this, \MANUAL\ did not outperform the optimization-based adaptation methods at any point, suggesting that user-driven adjustments based on preferences or intuition do not lead to better results.
This highlights the value of computational approaches over manual calibration for optimizing input devices.

Further analyzing regret values across user groups using \method, the first group (users 1–3), which was initially subjected to random explorations, had the highest cumulative regret.
On the other hand, later groups (users 7–9 and 10–12) outperformed the first, showing that \method\ benefits from prior user performance observations over time, leading to more accurate and efficient keyboard adaptations.

\section{Discussion}

In this work, we introduce CHiLO, a novel concept for boosting optimization efficiency by continually accumulating information across users. 
Based on our summarized principles, we further propose \method, a Bayesian optimization-based method that continually updates a population model to capture population-level characteristics. 
To further enhance \method’s robustness, we incorporate a memory-replay mechanism, which utilizes the previous users' models to train the population model. 
We evaluated \method through a user study focusing on a representative AR/VR input problem: mid-air typing on a virtual keyboard. 

The results showed that \method significantly outperforms two representative baselines --- standard BO and manual adjustment --- demonstrating its ability to effectively accumulate user experience and adapt more efficiently across different users. 
\method demonstrated a clear trend of learning even within 10 users: the earlier users exhibited higher regret than later ones, indicating that \method continually improves over time. 
Also, when using ConBO, only the first three users encountered randomly searched keyboard configurations. From the fourth user onward, all the generated keyboards are guided by the population model. 
This is a clear distinction to the previous meta-Bayesian optimization methods which require a fixed set of prior users that is significantly larger to go through a full optimization procedure.

In summary, ConBO shows positive results in gradually adapting more efficiently with an increasing number of users.
As Conbo is built upon BO, it does not impose strong assumptions or requirements for specific problems. We hope our results motivate researchers and practitioners to deploy ConBO for their applications. 
Despite these promising results, several open questions and limitations remain. 
There is potential for further optimizing the model’s scalability, refining its memory management, and exploring broader applications of CHiLO beyond the AR/VR contexts.

\subsection{Limitations and Future Work}

We believe our work makes a novel contribution by introducing CHiLO for dynamic user interface and interaction personalization, which we hope will inspire future research and development in the following areas:

\paragraph{Addressing Individual Diversity and User Sequence Challenges} 
\method leverages a memory-replay mechanism to ensure that the population model is trained on data from all prior users, minimizing the risk of catastrophic forgetting. 
However, a potential limitation is that the population model may primarily capture the characteristics of the ``general users,'' potentially leading to suboptimal performance for extreme users whose optimal settings significantly differ from the majority. 
This challenge is potentially severe in high-dimensional search spaces, where outliers could deviate even further from the average users.  
Future research should explore the use of more advanced models able to capture the diverse characteristics of different user groups. 
Additionally, the order in which users are encountered can influence the performance of \method. For instance, if the initial users represent average user behavior, the population model is more likely to transfer well to the next users. 
Conversely, if the first few users are outliers, the population model requires more iterations to adapt to the broader user population.  
This sequence effect is a known open challenge in continual learning \cite{ke2020continual, lopez2017gradient}. The extent of its impact depends on the nature of target interactions. Future research should empirically test the effects of user order in both synthetic tasks and real-world scenarios.

\paragraph{Re-adaptation to Changing User Characteristics}

Our current implementation assumes that user characteristics and optimal performance remain constant.
However, in practice, users' abilities may change due to factors like learning or fatigue, potentially requiring re-adaptation.
\method\ could handle these fluctuations by treating significant performance changes as signals to restart the adaptation process. For example, if a user's performance deviates significantly from the expected trajectory, the system could treat them as a new user and begin a new optimization cycle. Future work should focus on real-time detection of these changes and developing effective re-adaptation strategies.

\paragraph{Development of Alternative \concept Methods}
We see potential for further refinement of our proposed algorithm, as well as alternative implementations that may offer different trade-offs, making them more suitable for specific applications.
For instance, our current approach requires access to all prior user models, which may not be feasible for memory-constrained devices or environments without cloud access.
In such cases, a more size-efficient memory replay strategy could be necessary. 
A more refined approach could involve computing the importance of previous models and selectively using only the most relevant ones for population model training.

Also, our acquisition function is based on EI, a common option for BO. 
EI inherently balances exploration and exploitation, making it well-suited for general tasks. 
Future work could explore alternative acquisition functions, such as Probability of Improvement and Upper Confidence Bound, and empirically test their performances.
Also, instead of the greedy approach of selecting the design point with the highest acquisition value, one could investigate converting acquisition values into log-probabilities and sampling from them, further promoting exploration.
Our \method evaluates acquisition values by sampling over a fixed grid. Future work could explore more advanced optimization techniques, such as the DIRECT method \cite{jones1993lipschitzian} or gradient-based optimization with multiple restarts \cite{snoek2012practical}. These methods should improve the accuracy of finding the optimal acquisition value.

Additionally, while we update the population model after each user, exploring different update frequencies could help balance computational costs and model accuracy. Beyond using a BNN as the meta-surrogate, prior work has investigated meta-kernel approaches for continual learning in bandit optimization problems. Extending these kernel-based methods to broader optimization tasks presents another promising avenue for future research.

\paragraph{Continual Optimization across Applications and Potential Parallelization}
We introduce \method\ as a tool for improving adaptation by leveraging previous users' interactions with the same optimization task. However, we see the potential for \method\ to transfer knowledge not only across users but also across different applications, as many interactive systems share similar environments and parameter search spaces.
While our study focused on mid-air typing, future work should explore \method's generalizability across various input methods, such as touch typing, gesture-based input, and gaze-based input.
This would test \method's ability to learn shared characteristics across different interaction types and validate its broader applicability.
Advanced Bayesian Neural Network architectures, such as multi-task BNNs, could further enhance \method's capacity to encode both task-specific and shared information, enabling it to generalize effectively across users and applications.

Our population model is trained on all previously collected GPs, opening the possibility of parallel computing. The population model can be periodically updated on a centralized server while being duplicated and deployed to local devices. 
These local copies can be optimized for specific users or interactions in parallel. 
The updated GP models from local devices are then sent back to the server to jointly refine the population model, allowing for scalable and distributed optimization across multiple users and applications.

\paragraph{Expanding CHiLO for Complex Scenarios}
Our \method merely takes an initial step toward CHiLO, focusing on relatively simple single-objective problems. 
However, real-world HiLO scenarios are often more complex and may involve multiple objective functions, such as balancing accuracy and efficiency (e.g., \cite{10.1145/3491102.3501850}) or trading off between recognition and information transfer (e.g., \cite{10024515}). 
Addressing these challenges requires identifying the Pareto frontier within the objective space, a principled and widely used approach in multi-objective optimization tasks. 
Future work should explore extending CHiLO to handle multi-objective tasks by adapting the BNN to predict multiple objectives simultaneously and also updating the acquisition function to Expected Hypervolume Increase (EHVI), which is well-suited for identifying the multi-objective trade-offs.

Another common real-world scenario is where the objective functions cannot be explicitly measured, such as optimizing for user preferences.
Preferential Bayesian Optimization (PBO) addresses this issue by eliciting user preferences over a set of choices \cite{10.1145/3526113.3545664, 10.1145/3386569.3392444}. 
Extending CHiLO to preferential tasks is a direction worth exploring:
A preferential population model could be developed using a framework similar to \method, but additional research is needed to investigate methods for suggesting multiple design choices that leverage both the population model and the current user-specific model. 
Furthermore, empirical studies are essential to assess the practical effectiveness and scalability of CHiLO in PBO scenarios, particularly in capturing and adapting to diverse user preferences.

\paragraph{Enhancing Real-World Adaptation Efficiency}
Lastly, while \method improves the overall efficiency of keyboard adaptation, two areas remain for improvement:  (1) the first few users need to undergo random explorations, and (2) each user is required to type 28–32 characters for every keyboard update. 
Future HCI research should investigate alternative approaches to reduce such user efforts.  
A potential direction is incorporating established user models \cite{fischer2001user, oulasvirta2019s} to provide an initial estimate of general population performance. These models could replace or reduce the need for random exploration, enhancing the experience for the initial users.
Additionally, requiring 28–32 characters per optimization iteration for accurate performance estimation may not always be necessary. In some cases, a rough estimate (e.g., 10–20 characters) could suffice. 
Future research could investigate multi-fidelity Bayesian Optimization \cite{kandasamy2017multi, poloczek2017multi}, which accommodates evaluations of varying accuracy levels. Low-fidelity evaluations could quickly discard suboptimal designs, while high-fidelity evaluations could refine promising candidates.
Such advancements would make \method more efficient and user-friendly, extending its applicability beyond keyboard interaction and text input.

\section{Conclusion}

In this work, we address a novel challenge in Human in-the-loop Optimization: How can an optimizer continually accumulate experience and improve over time? The problem of continual learning for optimization, along with its technical challenges, had not been formally established in HCI, and through this work, we aim to bridge this gap by introducing CHiLO. We provide a formal mathematical framework and outline key design principles for its implementation.

Building on these principles, we propose \method, a Bayesian Neural Network-based approach that captures population-level user characteristics through a population model and incorporates a generative memory-replay mechanism using stored models from previous users. Our evaluation of \method on mid-air text input in VR demonstrated significant improvement over standard Bayesian optimization in terms of adaptation efficiency and overall performance, while offering performance comparable to manual adjustments without requiring explicit user feedback.

The results further show a clear trend of improvement over time, with later user groups benefiting from the accumulated knowledge of the system.
This illustrates \method's ability to progressively adapt more efficiently as it learns from prior users.
We hope these positive results encourage further research into applying \method to a wider range of applications, extending beyond HCI.

We believe that \method can be a key enabler for truly adaptive and personalized interaction in the future, with systems improving across users, devices, and tasks---ultimately evolving to enable efficient and frictionless interaction for a wide range of applications.

\section{Open Science}

\noindent
\method is available on our project page at \textit{\url{https://siplab.org/projects/Continual_Human-in-the-Loop_Optimization}}. We hope that our implementation encourages the development of more advanced methods and can help expand the application scope of CHiLO.

\vspace{4pt}

\begin{acks}
The authors sincerely thank Hee-Seung Moon, Thomas Langerak, Jiaxi Jiang, and Yufeng Zheng for their insightful discussions and support. We are grateful to Xintong Liu for her help on the figures. We also thank the participants of our user study.
Yi-Chi Liao was supported by the ETH Zurich Postdoctoral Fellowship Programme.
Paul Streli was in part supported by a Meta Research PhD Fellowship.
\end{acks}

\balance
\bibliographystyle{ACM-Reference-Format}
\bibliography{sample-base}
\appendix
\clearpage
\newpage

\section{Simulated Experiments Using Benchmark Functions}
\label{appendix:experiments}

The core idea of this paper is to develop a novel approach for training a population model to enable CHiLO. 
There are multiple potential ways to implement and train such a population model. In this section, we compare the optimization performance and computational efficiency of our ConBO approach (a BNN-based population model trained using predictions from previous models) against alternative approaches.
Specifically, our experiments aim to achieve the following goals:

\begin{itemize}[leftmargin=*]
    \item \textbf{Goal 1: population model performance comparison.} We compare different methods for constructing the population model and use them to sequentially optimize 15 user functions. Our goal is to demonstrate that ConBO's GP-based generative memory replay approach delivers superior optimization performance across users. 
    \item \textbf{Goal 2: ConBO with and without the user-specific model.} After confirming that ConBO’s population model performs as well or better than alternative methods, we examine the benefits of incorporating the user-specific model to compute acquisition values, showing potential improvements in performance, especially in later iterations. 
    \item \textbf{Goal 3: Computational efficiency.} We evaluate the computational cost of ConBO during deployment and compare it to other alternative implementations. The goal is to highlight the lightweight computational demands of ConBO, making it practical for real-world applications.
    \item \textbf{Goal 4: The effect of memory replay on forgetting .} Lastly, we analyze the effect of our memory-replay method on adaptation performance when revisiting previously encountered users.
\end{itemize}

In total, we run three simulated experiments. The first two experiments address the first three goals, while the last test focuses on the effect of memory replay.
Our results show that ConBO offers optimization performance comparable to or better than alternative approaches, with additional performance gains when the user-specific model is integrated. 
In terms of computational efficiency, we find that BNN-based population models like ConBO maintain relatively low computation durations.
Meanwhile, other GP-based population models experience increasing computation times as data grows, making them less suitable for direct application in CHiLO.
Moreover, we found that our memory replay effectively improves the optimization performance when encountering previously seen tasks, indicating an enhancement to reduce model forgetting. This indicates that our optimizer continuously improves over time, achieving greater efficiency for users with traits it has already seen in past optimizations.

For the following simulations, we use common benchmark functions to generate a set of user functions.
We begin by selecting a base function, such as the Branin function, to represent the population. By applying shifts and scaling transformations, we create a series of similar but distinct sub-functions, each representing an independent user. This allows us to simulate a scenario where the optimizer encounters different user-specific functions sequentially.
This approach is commonly used in simulation studies \cite{feurer2018practical, 10.1145/3613904.3642071}.

The following experiments were conducted on a Windows 10 system equipped with a 12th Gen Intel(R) Core i7 CPU and an NVIDIA GeForce RTX 3070 GPU.

\subsection{Different Approaches for Achieving CHiLO}
\label{appendix:different_approaches}

Here, we exhaustively outline different approaches for implementing and training a population model for CHiLO. 
To align with the broader BO framework, a population model must be capable of generating predicted means and variances for a given design candidate. 
These values are required for computing most acquisition functions in BO.
Two computational models that fulfill these requirements are GP and BNN, both of which can be trained on observations and produce the necessary predictions of means and variances.

First, we introduce the approaches using BNN as the core of the population model:

\paragraph{1. ConBO}
This is our proposed method, using a BNN as the population model. This BNN is continually retrained using the predicted means and variances generated from previous models, with each model representing a different user. 
Key strengths of ConBO include its well-distributed training data, which ensures the population model does not suffer from overfitting or underfitting in specific areas of the parameter space. Additionally, the BNN's predicted variance is carefully regulated: we filter out unreliable predictions based on variance thresholds, and the population model is trained to minimize predicted variance levels (see \autoref{loss}). This keeps the variance within a reasonable range, improving the model's reliability.
Furthermore, ConBO incorporates a GP model, representing the current user, to generate acquisition values, enabling more effective personalized adaptation and optimization. For a more detailed explanation of ConBO, please refer to \autoref{sec:conbo}.

\paragraph{2. ConBO without GP}
To assess the performance of ConBO's population model independently, we introduce a variation that excludes the influence of the current user's GP model, referred to as ``\emph{ConBO without GP}.'' This allows us to evaluate the population model's performance on its own.

\paragraph{3. ConBO without Variance Filter}
In the ``\emph{ConBO without Filter}'' approach, we omit the variance filter.
We assess the population model's performance without any constraints on the aggregated variances derived from prior user models.

\paragraph{4. BNN without Replay}
A simpler approach to using BNN as the population model is to continuously update it without incorporating any memory replay. This method resembles previous works in BO where large sample sizes necessitate the use of BNNs as surrogate models, allowing them to handle the growing computation time more efficiently than GPs (e.g., \cite{springenberg2016bayesian}). We call this approach \emph{BNN without Replay}.
To assess the performance of the population model independently, this approach also does not incorporate the current user's GP model. 
While straightforward to implement, this approach presents several challenges such as catastrophic forgetting and model instability due to the uneven distribution of observations (as mentioned in \autoref{sec:challenges_of_chilo}).
Furthermore, there is a potential for unstable variance. Without regulation, the BNN's predicted variance can become unstable (extremely high or low) as more data accumulates, complicating the computation of acquisition values.

\paragraph{5. BNN with Direct Replay}
Building on \textit{BNN without Replay}, a more advanced variation involves using memory replay. In this variation, the population model is updated with all observations from previous users after each user is completed. While this method effectively reduces forgetting, it faces potential challenges including unstable variances across the parameter space and the risk of overfitting in some regions while underfitting in others.
We refer to this variation as ``\textit{BNN with Direct Replay}.''

\vspace{0.25cm}
Below, we introduce the other approaches which leverage GP as the population model:

\paragraph{6. Single-GP}
Rather than using a BNN, one can employ a GP directly as the population model, capturing all observations from all users encountered. However, a major limitation of this approach is the cubic increase in computation time as more data is added. Beyond a certain point, updating the population model becomes prohibitively expensive and impractical for real-time applications.

\paragraph{7. Transfer Acquisition Function (TAF)}
To address the issue of cubic computation growth, previous works proposed a weighted-sum approach called Transfer Acquisition Function (TAF) \cite{wistuba2018scalable, 10.1145/3613904.3642071}, designed for meta-learning in BO. 
This approach stores each user's data as a separate GP model. When a new user is encountered, all previously stored models generate acquisition values, which are then combined based on weights to guide the optimization for the current user.
Typically, TAF has been applied in meta-learning scenarios where a fixed set of users, referred to as "prior users," undergo a complete optimization process to build a library of ``prior models.'' 
These models then aid in optimizing new users without adding further prior models.
This method can be extended to continual learning (CHiLO) by removing the distinction between ``prior users'' and ``new users'' In this setting, each new user's data is stored as an additional prior model, continuously growing the model library. Although this approach avoids the cubic increase in computation time, its computational cost still grows linearly with the number of models, which can become a bottleneck as the user base expands.

\paragraph{8. Standard BO (BO)}
Finally, we keep standard BO as a baseline, which utilizes the GP surrogate model to optimize for each current user.
There is no transferring knowledge between users.

\subsection{Base Function for Simulations}
In our simulations, we utilize two common base functions: Branin\footnote{\url{https://www.sfu.ca/~ssurjano/branin.html}} and McCormick\footnote{\url{https://www.sfu.ca/~ssurjano/mccorm.html}}, each with two design parameters and one objective function. They serve as the foundation for generating a group of similar yet distinct user functions.

We linearly normalize the input ranges of the Branin function ($x_1 \in [-5, 10], x_2 \in [0, 15]$) to match the McCormick function ($x_1, x_2 \in [0, 1]$), ensuring both have the same parameter range. 
While the output ranges of Branin and McCormick differ (Branin: approximately $[0.398, 300]$, McCormick: approximately $[-1.9133, 20]$), 
both functions were originally designed for minimization tasks.
We flip the functions (multiplying by $-1$) to convert them into maximization problems. 

\subsection{Generating a Group of Synthetic Users}

To simulate a diverse group of users with shared characteristics but varying behaviors, we create a set of sub-functions by shifting and scaling the base functions.
We will refer to these sub-functions as user functions.

Each user function is generated by shifting the input values before passing them to the base function. The magnitude of these shifts is sampled from a uniform distribution within a specified range, adding diversity among the users. Mathematically, the shift is represented as $x_n' = x_n + \delta_n$, where $x_n$ is the original input, $x_n'$ is the shifted input, $n \in {1, 2}$ represents the parameters, and the shift amount $\delta_n \sim U(-\frac{shift\_range}{2}, \frac{shift\_range}{2})$. This variation simulates different user responses to the same design.
We also scale the output of the base function by a scalar factor, which is also sampled from a uniform distribution. This factor, denoted as $\mathcal{S} \sim U(1-\frac{scale\_range}{2}, 1+\frac{scale\_range}{2})$, introduces further diversity by simulating users with different performance levels.

We first introduce the simulation experiment on Branin (A.4, A.5), followed by McCormick (A.6, A.7).

\subsection{Branin (Test 1): Task and Settings}
\label{appendix:branin}
We generate 15 user functions, with each optimization method running for 30 iterations per user function.  
For each user function, we perform a grid search to determine its optimal performance. Based on this, we compute the regret at each iteration. Additionally, we record the computation time at every iteration across all user functions and methods.
For this simulation, we set $shift\_range = 0.3$ and $scale\_range = 0.2$. Here, we provide the detailed settings for each approach:

\paragraph{1. ConBO}
The parameter settings for ConBO are aligned with those used in our user study.

\begin{itemize}[leftmargin=*]
  \item The \emph{population model} is a BNN consisting of three fully connected layers, each with 100 nodes using ReLU as the activation function. A dropout layer (dropout rate of 0.1) is applied after the last hidden layer. During population model training (Step 4 in the workflow), the BNN is trained for 1200 epochs. For adaptation (Step 1), we run 15 epochs for each new observation.
  \item The \emph{user-specific models} (current and prior) are implemented as Gaussian Processes (GPs) with a Matérn $5/2$ kernel, accounting for homoscedastic noise.
  \item The initial number of random exploration points, $r_0$, is set to 16, with a decay rate $d_r$ of 5. We merge the EI values from both the population model and the current user-specific model using a weighting factor from Eq.~\ref{eq:weighting}, where $\alpha_1 = 16$ and $\alpha_2 = 0.1$. This ensures that the influence of the current user’s model increases after the fifth iteration, growing linearly.
  \item For population model training, we first generate a grid of $30 \times 30$ points evenly spaced across the design space, followed by 50 randomly sampled points. The predicted means and variances from all previous user-specific models are then queried. A variance threshold of $\lambda = 50$ is applied.
\end{itemize}

\paragraph{2. ConBO without GP}
This variant's setting is similar to ConBO, except for setting $\alpha_1 = 30$, meaning that all EI values are derived from the population model, as no adaptation from the user-specific model is applied.

\paragraph{3. ConBO without Variance Filter}
This variant is similar to ConBO, with the key difference being the disabled variance filter. Consequently, all mean and variance estimates across the design space from every previous model are incorporated into the training of the population model. This includes estimates from a user's model in regions where it has encountered few or no observations---an inclusion that the variance filter would otherwise exclude.

\paragraph{4. BNN without Replay}
All other settings are identical to those in ConBO, except that there is no memory replay to retrain the population model. Instead, the population model keeps updating based on new observations.

\paragraph{5. BNN with Direct Replay}

Similar to the \textit{BNN without Replay}, but here, after each user’s optimization is complete, the population model is retrained on all past observations.

\paragraph{6. Single GP}

This method uses a single-task GP with a Matérn $5/2$ kernel. The optimization process involves 10 restarts for the acquisition function, with 1,024 candidate points and 512 Monte Carlo samples. Of the 30 total iterations, the first 16 are random searches, followed by 14 optimization iterations guided by the Expected Improvement acquisition function.
We retain the GP as a population model and update it with new observations over time. 
The number of random search decay across users, with an initial $r_0$ of 16 and a decay rate $d_r$ of 5.

\paragraph{7. Transfer Acquisition Function (TAF)}

The TAF setting mostly follows prior work \cite{10.1145/3613904.3642071}, where observations from each user function are stored in independent GP models. 
The number of random searches decays across users, with an initial $r_0$ of 16 and a decay rate $d_r$ of 5. 
We further follow the original paper's settings, with weights decaying using $d_1 = 0$ and $d_2 = 0.05$, which were identified through pilot tests to enhance performance.

\paragraph{8. Standard BO (BO)}

This approach basically follows \textit{Single GP} but the GP model restarts for each new user.

\subsection{Branin (Test 1): Results}

\begin{figure*}[t!]
    \centering
    \includegraphics[width=\linewidth]{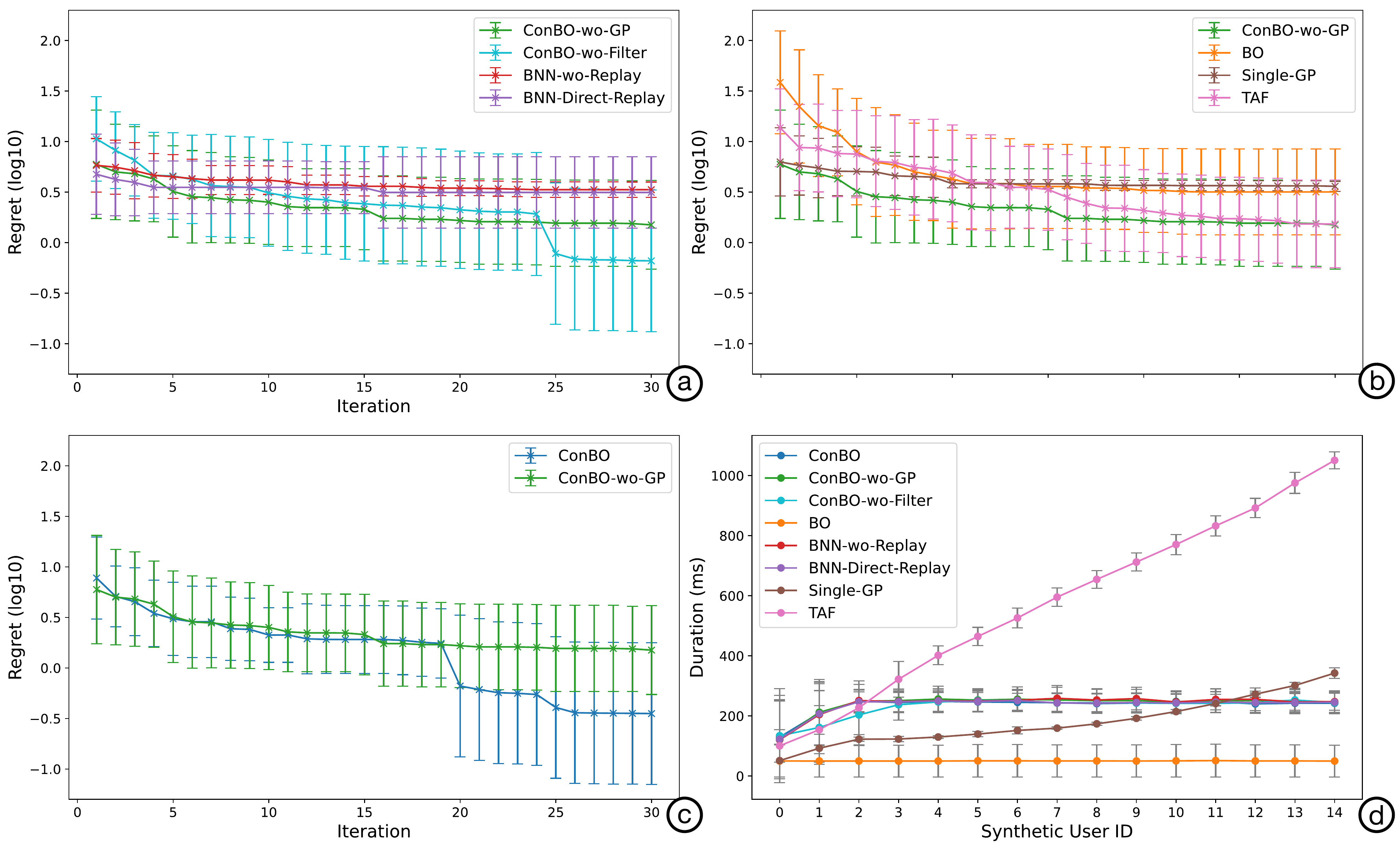}
    \caption{Results of our simulation study using the Branin function (Test 1). All error bars represent one standard deviation. (a) The regret values over the iterations of different BNN-based methods. We highlight that our \textit{ConBO without GP} has better performance over other BNN-based population models. (b) The regret values over the iterations of GP-based methods, highlighting \textit{ConBO without GP} outperforms Single GP and delivers comparable performance to TAF. (c) The regret values over the iterations to compare ConBO with and without integrating the user-specific model. The result shows that, with the support of the user-specific model, ConBO can improve its performance further. (d) The mean computation time spent in one iteration for each user. The result highlights the computation costs of GP-based population models (Single GP and TAF) increase quickly when accumulating more data.}
    \label{fig:sim1}
\end{figure*}

\autoref{fig:sim1} shows the results of our simulation on the Branin function.

\begin{figure*}[t!]
    \centering
    \includegraphics[width=\linewidth]{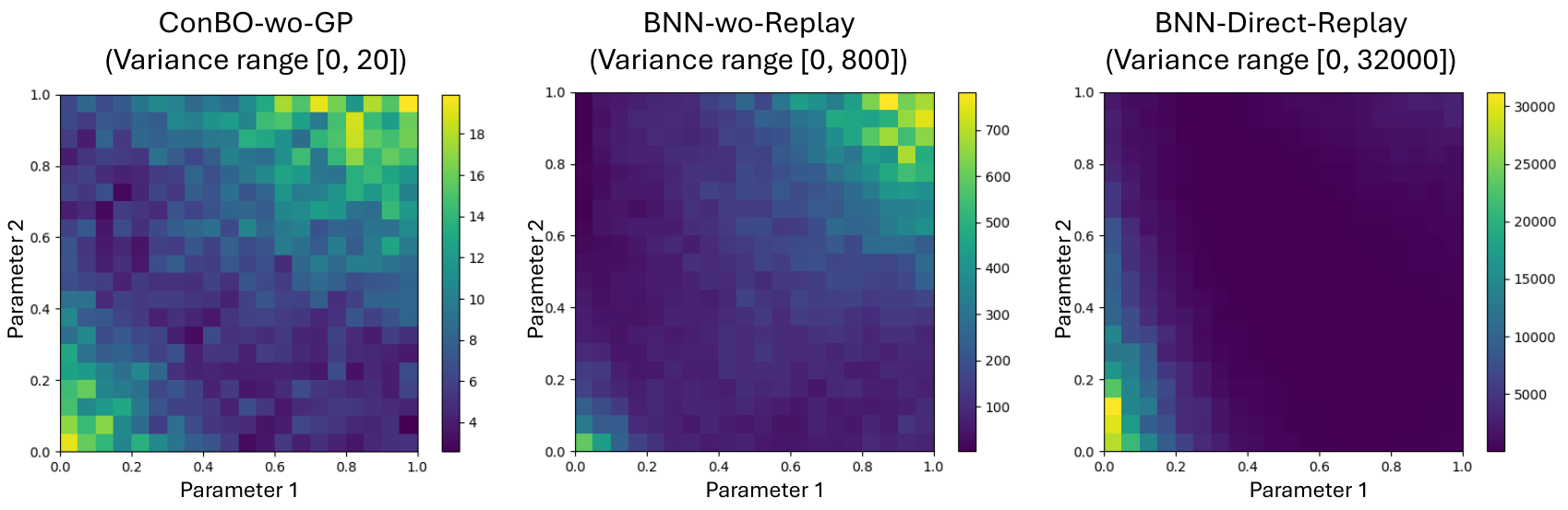}
    \caption{The predicted variance values derived from different BNN-based population model implementations after all 15 user functions. Our ConBO utilizes previous GPs to generate predicted means and variances and further utilizes a variance filter to remove unreliable predictions. Therefore, ConBO allows for more regulated variance predictions. On the other hand, other BNN-based approaches do not have an explicit mechanism to regulate the variance of the population model, making it unstable and changing drastically over the parameter space when trained with a large number of observations. This highlights the stability offered by our approach.} 
    \label{fig:sim1_var}
\end{figure*}

\paragraph{\textbf{Goal 1: population model performance comparison}}
To analyze the performance of our approach with respect to our first goal, we compare \textit{ConBO without GP} against other approaches (\autoref{fig:sim1}a and \autoref{fig:sim1}b). 
Due to the large number of baselines, we separate the approaches into BNN-based (\autoref{fig:sim1}a) and GP-based (\autoref{fig:sim1}b) groups.
Examining \autoref{fig:sim1}a, we found that \textit{ConBO without GP} surpasses other BNN-based variations. Although these methods show comparable performance in the early iterations, they fail to adapt effectively to specific functions over time, resulting in poorer performance in later iterations.
Further analysis (\autoref{fig:sim1_var}) shows a key issue with the direct replay and no-replay approaches: the predicted variances in the BNN models are unregulated. 
As more data is introduced, these methods exhibit highly unstable variance estimates, with values becoming excessively high.
Moreover, since the training data is unevenly distributed across the parameter space, the predicted variances vary significantly. 
For instance, in the BNN without replay, variance values range from under 100 to over 700.
Given that variance plays a critical role in computing the acquisition function, this instability ultimately hinders the models' ability to adapt effectively. 
In contrast, ConBO employs a grid-based generative approach with a variance filter, ensuring that the training data for the population model remains within a reasonable range and spans across the parameter space evenly. 
This regulation leads to more stable variance predictions and, consequently, better online adaptation performance.
Finally, we found that ConBO-wo-GP performed better than ConBO-wo-Filter in early iterations, owing to the more regulated variance range.

Comparing \textit{ConBO without GP} against GP-based approaches (\autoref{fig:sim1}b), \textit{ConBO without GP} outperforms standard BO, which typically starts optimization from scratch, leading to slower adaptation and worse performance in early iterations. This is reflected in its higher regret values, particularly in the initial stages of optimization.
In addition, both \textit{ConBO without GP} and TAF outperform the Single GP approach. This is also expected, as the single GP model lacks variance regulation, potentially leading to unstable model fitting when encountering a large number of observations and ineffective user-specific adaptation.

\paragraph{\textbf{Goal 2: ConBO with and without the user-specific model}}
To address our second goal, we compare the performance of ConBO with and without the current user's GP model in \autoref{fig:sim1}c.
It demonstrates that incorporating the user-specific model to generate acquisition values allows ConBO to further enhance its performance, particularly in later iterations. This highlights the importance of leveraging a user-specific model to support the adaptation process, leading to more personalized results.

\paragraph{\textbf{Goal 3: Computational efficiency}}
To answer our third goal, we show the average computation duration at each iteration across all users in \autoref{fig:sim1}d. 
This highlights a significant drawback of GP-based approaches. 
The computation time of a single GP increases cubically with the number of observations.
Specifically, the computation time of a single GP surpasses that of BNN-based approaches around the 11th or 12th synthetic user, corresponding to approximately 350 observations. This number is easily reached in real-world deployments involving a sequence of actual users.
On the other hand, TAF's computation time grows linearly with the number of prior models (i.e., previously encountered users). 
Notably, the computation time of TAF surpasses the BNN-based methods from the 4th synthetic user onward.
This increase in computation cost makes the GP-based solutions unsuitable for scenarios with a potentially unlimited number of users in a real-world continual learning setting.
In contrast, the BNN-based approaches, including ConBO, do not suffer from the increasing computation duration.

\subsection{McCormick (Test 2): Task and Settings}

As with the Branin simulation, we generate 15 user functions for the McCormick function. As McCormick is a simpler function, each user function is optimized over 7 iterations. 
For each user function, we perform a grid search to determine its optimal performance. Based on this, we compute the regret at each iteration. Additionally, we record the computation time at each iteration across all user functions and methods. 
In this simulation, we set the shift range to \(shift\_range = 0.5 \) and the scale range to \(scale\_range = 0.2 \).

Below are the detailed settings for each approach:

\paragraph{1. ConBO}
The configuration of ConBO is similar to the one used in the Branin function (Test 1). 
However, due to the limited optimization budget of 7 iterations, the initial number of random exploration points, \( r_0 \), is set to 3, with a decay rate \( d_r \) of 3. We combine the EI values from both the population model and the user-specific model, with weighting factors \( \alpha_1 = 3 \) and \( \alpha_2 = 0.15 \). Additionally, a variance threshold of \( \lambda = 10 \) is applied.

\paragraph{2. ConBO without GP}
This configuration mirrors ConBO, with the exception that \( \alpha_1 \) is set to 7, meaning all EI values are derived exclusively from the population model, without incorporating the user-specific model.

\paragraph{3. ConBO without Variance Filter}
This variant's setting is similar to ConBO, except for disabling the variance filter, meaning that all the estimates across the design space from every previous user model are used for population model training.

\paragraph{4. BNN without Replay}
This approach follows a setup similar to the Branin simulation, with the only difference being the number of random exploration points. Here, the initial \( r_0 \) is set to 3, with a decay rate \( d_r \) of 3.

\paragraph{5. BNN with Direct Replay}
This also mirrors the its setting in the Branin simulation but differs in the number of random exploration points. Here, the initial \( r_0 \) is set to 3, with a decay rate \( d_r \) of 3.

\paragraph{6. Single GP}
This approach's setting is similar to one used in the Branin simulation, with the only difference being the number of random exploration points. Here, the initial \( r_0 \) is set to 3, with a decay rate \( d_r \) of 3.

\paragraph{7. Transfer Acquisition Function (TAF)}
This approach shares similar settings as in the Branin simulation, where \( r_0 = 3 \) and the decay rate \( d_r  = 3\). Additionally, the weight decay parameters for the prior models are consistent with the Branin setup: \( d_1 = 0 \) and \( d_2 = 0.15 \).

\paragraph{8. Standard BO}
For the 7 total iterations, the first 3 iterations are random searches, followed by 4 optimization iterations guided by the Expected Improvement acquisition function.

\subsection{McCormick (Test 2): Results}
\begin{figure*}[t!]
    \centering
    \includegraphics[width=\linewidth]{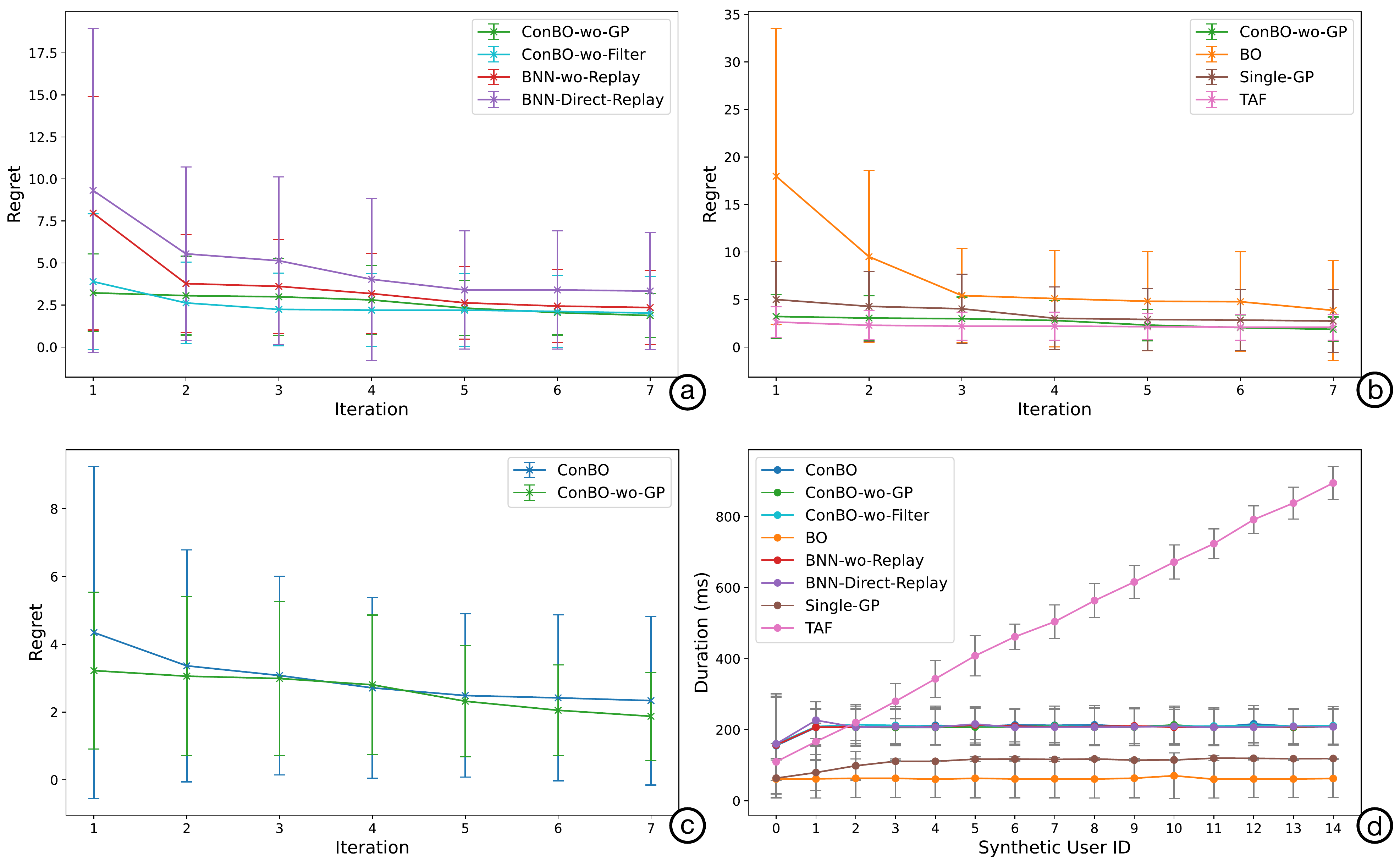}
    \caption{Results of our simulated study using the McCormick function (Test 2). All error bars represent one standard deviation. (a) The regret values over the iterations of different BNN-based methods. We highlight that our \emph{ConBO without GP} has better performance over other BNN-based population models. (b) The regret values over the iterations of different GP-based methods. \textit{ConBO without GP} outperforms standard BO overall and Single GP in early iterations. \textit{ConBO without GP} and TAF provide similar performance.
    (c)  The regret values over the iterations to compare ConBO with and without integrating the user-specific model. Incorporating the user-specific model does not lead to a significant difference in performance.
    (d) The mean computation time spent in one iteration for each user. The result highlights that the computation time of TAF increases linearly with the number of users.}
    \label{fig:sim2}
\end{figure*}

\autoref{fig:sim2} presents the results of the McCormick simulation, which align with our findings from the Branin function.

\paragraph{\textbf{Goal 1: population model performance comparison}}
To determine whether ConBo meets our first goal in this setting, we compare \textit{ConBO without GP} to other approaches (\autoref{fig:sim2}, a and b). 
Again, we separate approaches into BNN-based (shown in \autoref{fig:sim2}, a) and GP-based (shown in \autoref{fig:sim1}, b) groups.
In \autoref{fig:sim2} (a), while all BNN-based methods converge by the final iteration, we observe that \textit{ConBO without GP} and ConBO without Variance Filter outperform other BNN-based variations in the earlier iterations.
This is consistent with our findings from the Branin function.
In \autoref{fig:sim2} (b), we see that \textit{ConBO without GP} outperforms standard BO overall and Single GP in early iterations. Both \textit{ConBO without GP} and TAF provide similar performance throughout the simulation.

\paragraph{\textbf{Goal 2: ConBO with and without the user-specific model}}
To answer our second goal, \autoref{fig:sim2} (c) compares the two variations of ConBo. In this case, incorporating the user-specific model does not lead to a significant difference in performance.
This can potentially be attributed to the lower complexity of the McCormick function compared to the Branin function. As a result, variations in the user functions do not impact performance to the same extent.

\paragraph{\textbf{Goal 3: Computational efficiency}}
To address our third goal, \autoref{fig:sim2} (d) shows the computation time per iteration across users. As expected, TAF shows a linearly increasing computation time. 
As observed in the previous simulated test, its computation time surpasses that of the BNN-based methods starting from the 4th synthetic user.
Notably, since each user in the McCormick simulation has significantly fewer observations than in the Branin simulation (7 vs. 30 iterations for each user), the Single GP approach has not yet reached the point of rapidly escalating computational cost, as seen in \autoref{fig:sim1}d. 
However, if the synthetic users continued, the computation time for a single GP would surpass that of BNN-based methods at approximately 350 observations (around 50 synthetic users) as we learned from the previous simulated test.

\subsection{Revisiting Previously Seen Synthetic Users (Test 3)}
\label{appendix:forget}

\subsubsection{Task and Settings}
In Test 3, we aim to demonstrate how \method's memory-replay mechanism addresses the potential forgetting issue in CHiLO. 
To focus specifically on the impact of the memory-replay mechanism, we compare two approaches: \emph{1. ConBO (with Replay)} and \emph{2. BNN without Replay}. 
The experimental settings for these two approaches are identical to those described in Test 1 (Appendix \autoref{appendix:branin}).

We use the Branin function as the base to generate 10 synthetic users, with a shift range of 0.4 and a scale range of 0.4. These ranges are slightly broader than those in Test 1, resulting in a more diverse set of synthetic users.

This test is divided into two phases: \emph{initial optimization} and \emph{re-optimization}.
In initial optimization, both approaches continually optimize for the 10 synthetic users, and the resulting population models are stored as the \textbf{trained models}.
In the subsequent re-optimization phase, these two \textbf{trained models} are used to re-optimize the same previously encountered synthetic users. This evaluates whether the population models effectively retain the memory of previously seen tasks and whether the optimizer improves its efficiency for users with characteristics similar to those it has seen before.
Note that during re-optimization, we employ the same trained population models for each synthetic user (i.e., the re-optimization of individual users does not affect the model for other users).
These trained population models capture the state after \textit{initial optimization}.
This analysis evaluates the ability of the population models to retain knowledge about prior users after all synthetic users in the initial optimization phase.

\subsubsection{\textbf{Goal 4: The effect of memory replay on forgetting}}
\begin{figure*}[t!]
    \centering
    \includegraphics[width=\linewidth]{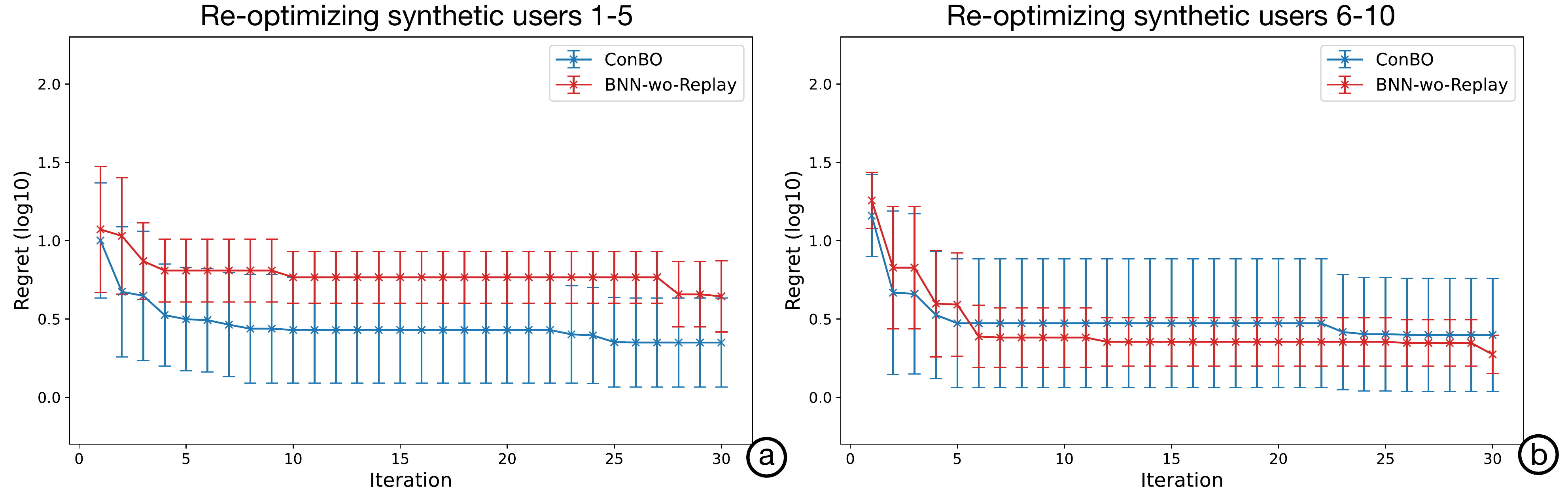}
    \caption{Results of our simulated test when \emph{re-optimizing for the previously seen users} (Test 3). All error bars represent one standard deviation. (a) The performance for re-optimizing the 1st--5th synthetic users. (b) The performance for re-optimizing the 6th--10th synthetic users.}
    \label{fig:sim3}
\end{figure*}

The results of the re-optimization are shown in \autoref{fig:sim3}.
For the re-optimization of users 1 to 5, the population model trained with \textit{BNN without Replay} performs significantly worse compared to the model trained with \textit{ConBO}.
This suggests that, without a memory-replay mechanism, the population model struggles to retain the knowledge of earlier users, leading to reduced optimization efficiency when revisiting them. In comparison, for the more recent 6th to 10th synthetic users, the performance of \textit{BNN without Replay} is comparable to that of \textit{ConBO}.

\subsection{Findings and Conclusion}

Through the simulations, we explored the effectiveness of our proposed ConBO method compared to alternative approaches for constructing a population model for CHiLO. 
Particularly, our experiments on the Branin and McCormick functions aimed to address three key goals: comparing population models' optimization performance, understanding the benefits of incorporating a user-specific model, and evaluating computational efficiency.

Our results show that ConBO outperforms other BNN-based approaches even when the user-specific model is not included, thanks to its regulated variance predictions and well-distributed training data across the parameter space. This ensures that ConBO avoids the pitfalls of overfitting or underfitting, providing consistent performance across different users. When the user-specific model is incorporated, ConBO demonstrates further improvements in performance, particularly in later iterations, allowing for more personalized and adaptive optimization.

In terms of computational efficiency, ConBO maintains a lightweight computational profile, even as more data is accumulated. This stands in contrast to other GP-based approaches, such as Single GP and TAF, which experience significant computational growth as the dataset scales, making them less suitable for continual learning applications.
Furthermore, ConBO (with memory replay) showcases a faster convergence when re-optimizing the previously seen synthetic users. This highlights its ability to address the potential forgetting issue in CHiLO.

Overall, ConBO offers an effective balance between optimization performance and computational efficiency, making it a strong candidate for population model construction in CHiLO applications.

\section{Analysis of User Diversity}
\label{appendix:user_diversity}

This section provides a detailed analysis of the diverse user characteristics to illustrate how ConBO effectively adapts to varying user performances.
Specifically, we examine the optimal keyboard design for each individual user and their corresponding typing performance.
Additionally, we show how the population model evolves over the sequence of users.

\subsection{Participants Characteristics}
\label{appendix:user_details}

The details of all participants are provided in \autoref{tab:users}, showcasing the diverse range of user characteristics in our user study. Note that all the participants had normal eyesight or corrected eyesight (glasses or contact lenses) during the study.

\begin{table*}[ht]
\centering
\resizebox{1\textwidth}{!}{
\begin{tabular}{|c|c|c|c|c|c|c|c|}
\hline
\textbf{Participant ID} & \textbf{Age} & \textbf{Gender} & \textbf{Height (cm)} & \textbf{Hand Length (cm)} & \textbf{Hand Breadth (cm)} & \textbf{Arm Length (cm)} & \textbf{Eyesight} \\ \hline
1 & 26 & Male   & 178 & 17.9 & 7.0 & 60 & Corrected \\ \hline 
2 & 38 & Male   & 190 & 20.5 & 8 & 66 & Normal \\ \hline 
3 & 26 & Male   & 186 & 19 & 7.5 & 60 & Normal \\ \hline 
4 & 30 & Male  & 180 & 18.5 & 7.1 & 59 & Corrected \\ \hline 
5 & 27 & Male   & 183 & 18.7 & 7.8 & 61 & Normal \\ \hline 
6 & 25 & Female & 160 & 16.5 & 6 & 47 & Corrected \\ \hline 
7 & 26 & Male   & 192 & 20 & 8 & 70 & Normal \\ \hline 
8 & 28 & Male   & 195 & 20.5 & 8.1 & 69 & Normal \\ \hline 
9 & 30 & Male  & 169 & 17.2 & 6.3 & 54 & Corrected \\ \hline 
10 & 24 & Male   & 185 & 17.9 & 6.9 & 65 & Normal \\ \hline 
11 & 25 & Male  & 175 & 17 & 6.4 & 57 & Corrected \\ \hline %
12 & 28 & Female   & 169 & 16.3 & 6.3 & 53 & Corrected \\ \hline 
\end{tabular}
}
\caption{Participant characteristics. }
\label{tab:users}
\end{table*}

\subsection{Optimal Keyboard Design for each User}
\label{appendix:optimal_keyboards}

\autoref{tab:merged_summary} presents the optimal keyboard configuration and corresponding typing performance for each user under the three adaptation conditions \MANUAL, \STANBO, and \CONBO, according to our text entry study.
Particularly, both \CONBO\ and \STANBO\ identified a higher variety of optimal keyboard dimensions (from $32.5 \times 11.37$ to $88.12 \times 30.84$). However, \MANUAL\ leads to less diverse keyboard dimension setting.

Our analysis shows that \MANUAL\ exhibits less variation in keyboard configuration (std. width: 6.69\,cm, distance: 10.84\,cm) compared to \STANBO\ (std. width: 19.88\,cm, distance: 12.87\,cm) and \CONBO\ (std. width: 16.25\,cm, distance: 14.85\,cm), suggesting that users tend to be more conservative with their manual keyboard setup.
Moreover, \CONBO\ demonstrates the ability to remain adaptive to individual users with a growing user base.

\begin{table*}[htbp]
    \centering
    \resizebox{1\textwidth}{!}{
    \begin{tabular}{|c|ccc|ccc|ccc|}
        \hline
        \textbf{Pariticipant ID} & \multicolumn{3}{c|}{\textbf{Width, Height [cm]}} & \multicolumn{3}{c|}{\textbf{Distance [cm]}} & \multicolumn{3}{c|}{\textbf{Typing Speed [NetWPM]}} \\
         & \textbf{MANUAL} & \textbf{STANBO} & \textbf{CONBO} & \textbf{MANUAL} & \textbf{STANBO} & \textbf{CONBO} & \textbf{MANUAL} & \textbf{STANBO} & \textbf{CONBO} \\
        \hline
        1  & 40.00, 14.00  & 37.5, 13.12 & 32.50, 11.37 & 49.21 &  29.00& 43.67 & 25.00 & 16.65 & 24.76 \\
        2  & 50.00, 17.50  & 88.12, 30.84 & 88.12, 30.84 & 28.16 & 64.00 & 36.00 & 19.75 & 15.87 & 16.42 \\
        3  & 40.00, 14.00  & 58.13, 20.34 & 86.25, 30.19 & 54.47 & 53.00 & 31.00 & 10.24 & 13.03 & 12.47 \\
        4  & 50.00, 17.50  & 78.75, 27.56 & 76.87, 26.91 & 49.21 & 50.00 & 31.00 & 18.99 & 19.96 & 19.18 \\
        5  & 50.00, 17.50  & 20.62, 7.22  & 88.12, 30.84 & 28.16 & 41.00 & 38.00 & 17.45 & 9.63  & 14.02 \\
        6  & 60.00, 21.00  & 71.25, 24.94 & 82.50, 28.88 & 61.84 & 64.00 & 25.00 & 16.51 & 16.77 & 22.94 \\
        7  & 50.00, 17.50  & 61.88, 21.66 & 88.12, 30.84 & 43.95 & 47.00 & 31.00 & 15.07 & 14.27 & 13.12 \\
        8  & 60.00, 21.00  & 54.38, 19.03 & 76.87, 26.91 & 38.68 & 44.00 & 64.00 & 13.42 & 15.13 & 14.94 \\
        9  & 50.00, 17.50  & 31.87, 11.16 & 88.12, 30.84 & 38.68 & 37.00 & 63.00 & 12.76 & 14.25 & 14.28 \\
        10 & 50.00, 17.50  & 75.00, 26.25 & 75.00, 26.25 & 59.74 & 26.00 & 64.00 & 20.88 & 20.38 & 22.72 \\
        11 & 60.00, 21.00  & 58.13, 20.34 & 86.25, 30.19 & 43.95 & 57.00 & 59.00 & 13.56 & 12.12 & 11.31 \\
        12 & 50.00, 17.50  & 61.88, 21.66 & 61.88, 21.66 & 49.21 & 61.00 & 53.00 & 9.56  & 17.47 & 16.24 \\
        \hline
    \end{tabular}
    }
    \caption{Summary of keyboard dimensions, distance, and typing speed for each user across MANUAL, STANBO, and CONBO.}
    \label{tab:merged_summary}
\end{table*}

\subsection{Evolvement of the population model over Participants}
\label{appendix:meta_model}

ConBO's population model captures prior beliefs about expected performance based on the observed performance of previous participants. 
\autoref{tab:range} shows the mean, minimum and maximum parameter values for the optimal keyboard configurations over the first 3, 6, 9, and 12 participants.

\begin{figure*}[t!]
    \centering
    \includegraphics[width=1\linewidth]{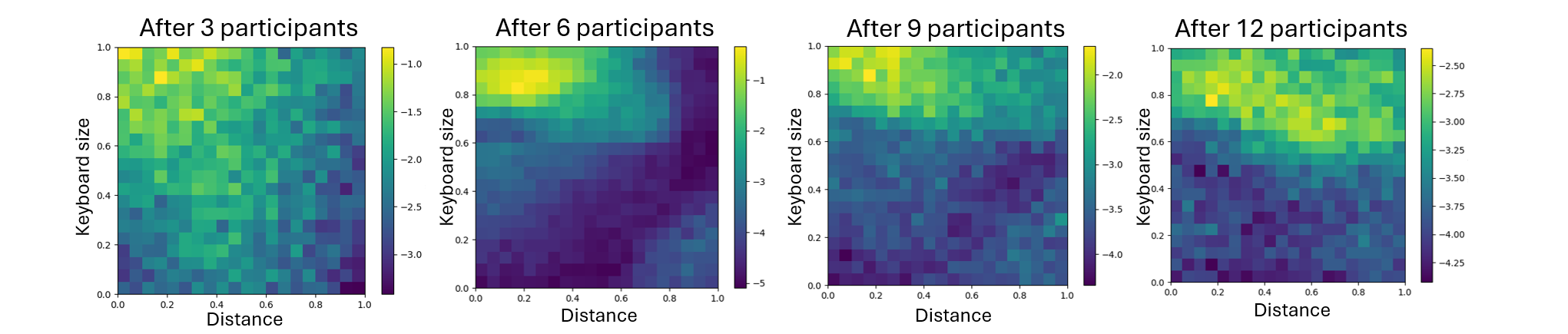}
    \caption{Estimated typing performance for different keyboard configurations according to the BNN population model with a growing user base. Note that the design parameters are normalized to the range $[0, 1]$, while predicted performance is normalized to $[-5, 5]$.}
    \label{fig:bnn_prediction}
\end{figure*}

In \autoref{fig:bnn_prediction}, we observe that the population model’s predicted performance evolves in alignment with the observed user performance.
Initially, after collecting data from the first three participants, the population model lacks sufficient information, resulting in a relatively uniform prior over the possible keyboard configurations. An early pattern emerges, suggesting that larger keyboard sizes tend to improve performance.
After incorporating data from six participants, this pattern becomes more pronounced, with larger keyboard sizes showing improved performance. At the same time, the population model suggests that shorter distances between the user and the keyboard can enhance typing performance.
By the time data from 12 participants is added, the population model adjusts, recognizing that some users perform better with the keyboard placed farther away and that the ideal distance may vary from person to person, while still generally favoring larger keyboard sizes.

This gradual learning process demonstrates the population model's ability to refine its estimates as it continuously adapts to new participant data, identifying performance patterns across previous users.
The population model's improving prior belief about expected user performance leads to increasingly smaller regret with growing numbers of participants (see \autoref{fig:result}).

\begin{table*}[htbp]
\centering
\resizebox{1\textwidth}{!}{
\begin{tabular}{|l|ccc|ccc|ccc|}
    \hline
    {\textbf{Group}} & \multicolumn{3}{c|}{\textbf{Width [cm] (normalized)}} & \multicolumn{3}{c|}{\textbf{Height [cm] (normalized)}} & \multicolumn{3}{c|}{\textbf{Distance [cm] (normalized)}} \\
    & \textbf{Mean} & \textbf{Min} & \textbf{Max} & \textbf{Mean} & \textbf{Min} & \textbf{Max} & \textbf{Mean} & \textbf{Min} & \textbf{Max} \\
    \hline
    P 1-3  &  68.96 (0.71) &  32.5 (0.23) &  88.12 (0.97) &  24.13 (0.71) &  11.37 (0.23) &  30.84 (0.97) &   36.89 (0.29) &   31.0 (0.15) &  43.67 (0.46) \\
P 1-6  &  75.73 (0.80) &  32.5 (0.23) &  88.12 (0.97) &  26.51 (0.80) &  11.37 (0.23) &  30.84 (0.97) &   34.11 (0.22) &    25.0 (0.00) &  43.67 (0.46) \\
P 1-9  &  78.61 (0.84) &  32.5 (0.23) &  88.12 (0.97) &  27.51 (0.84) &  11.37 (0.23) &  30.84 (0.97) &    40.3 (0.38) &    25.0 (0.00) &   64.0 (0.97) \\
P 1-12 &  77.55 (0.83) &  32.5 (0.23) &  88.12 (0.97) &  27.14 (0.83) &  11.37 (0.23) &  30.84 (0.97) &   44.89 (0.50) &    25.0 (0.00) &   64.0 (0.98) \\

    \hline
\end{tabular}
}
\caption{Mean, minimum, and maximum of the optimal keyboard configurations for all participants and subgroups (normalized values according to \autoref{fig:bnn_prediction}).}
\label{tab:range}
\end{table*}

\end{document}